\documentclass[11pt]{article}
\usepackage{geometry}               
\usepackage{cite}
\usepackage[dvipdfmx]{graphicx,color,hyperref}
\hypersetup{colorlinks=true, citecolor=, linkcolor=, urlcolor=}
\usepackage{amssymb,amsmath,bm,braket}
\makeatletter

\@addtoreset{equation}{section}
\makeatother
\usepackage{authblk}
\usepackage{arydshln}

\newcommand{\susic}{supersymmetric }
\begin{document}
%\title{}
%\author[]{}
%\date{}                                      

%%%%%%%%%%%%%%%%%%%%%%%%%%%%%%%%%%%%%%%%%%%%%%%%%%%%%%%%%%

\newcommand{\Om}{\Omega}
\newcommand{\om}{\omega}
\newcommand{\al}{\alpha}
\newcommand{\ep}{\epsilon}
\newcommand{\de}{\delta}
\newcommand{\la}{\lambda}
\newcommand{\La}{\Lambda}

%%%%%%%%%%%%%%%%%%%%%%%%%%%%%%%%%%%%%%%%%%%%%%%%%%%%%%%

\newcommand{\deebar}{\bar{\partial}}

\newcommand{\del}{\partial}

\newcommand{\df}{\stackrel{\rm def}{=}}
\newcommand{\co}{{\scriptstyle \circ}}

\newcommand{\lb}{\lbrack}
\newcommand{\rb}{\rbrack}
\newcommand{\rn}[1]{\romannumeral #1}
\newcommand{\msc}[1]{\mbox{\scriptsize #1}}
\newcommand{\dsp}{\displaystyle}
\newcommand{\scs}[1]{{\scriptstyle #1}}

\newcommand{\bc}{\Bbb C}
\newcommand{\br}{\Bbb R}
\newcommand{\bz}{\Bbb Z}
\newcommand{\bq}{\Bbb Q}
\newcommand{\bn}{\Bbb N}
\newcommand{\bs}{\Bbb S}
\newcommand{\bh}{\Bbb H}

\newcommand{\bsc}{\Bbb C}
\newcommand{\bsr}{\Bbb R}
\newcommand{\bsz}{\Bbb Z}

\newcommand{\ba}{\mbox{{\bf a}}}
\newcommand{\bsa}{\msc{{\bf a}}}
\newcommand{\da}{\dot{a}}
\newcommand{\dal}{\dot{\alpha}}
\newcommand{\db}{\dot{b}}
\newcommand{\dbeta}{\dot{\beta}}

\newcommand{\g}{\mbox{{\bf g}}}
\newcommand{\f}{\mbox{{\bf f}}}
\newcommand{\G}{\mbox{{\bf G}}}
\newcommand{\F}{\mbox{{\bf F}}}

\renewcommand{\-}{{\bf -1}}
\newcommand{\sigmab}{\sigma_{\msc{b}}}
\newcommand{\sigmaf}{\sigma_{\msc{f}}}

\newcommand{\cA}{{\cal A}}
\newcommand{\cL}{{\cal L}}
\newcommand{\cG}{{\cal G}}
\newcommand{\cJ}{{\cal J}}
\newcommand{\cT}{{\cal T}}
\newcommand{\cO}{{\cal O}}
\newcommand{\cN}{{\cal N}}
\newcommand{\cM}{{\cal M}}
\newcommand{\cF}{{\cal F}}
\newcommand{\cP}{{\cal P}}
\newcommand{\cS}{{\cal S}}
\newcommand{\cR}{{\cal R}}
\newcommand{\cC}{{\cal C}}
\newcommand{\cQ}{{\cal Q}}
\newcommand{\cD}{{\cal D}}
\newcommand{\cE}{{\cal E}}
\newcommand{\cH}{{\cal H}}
\newcommand{\cU}{{\cal U}}
\newcommand{\cZ}{{\cal Z}}
\newcommand{\cI}{{\cal I}}
\newcommand{\cK}{{\cal K}}

\newcommand{\wcH}{\widehat{\cal H}}
\newcommand{\wcN}{\widehat{\cal N}}

\newcommand{\wtR}{\widetilde{R}}

\newcommand{\tL}{\tilde{L}}
\newcommand{\tJ}{\tilde{J}}
\newcommand{\tI}{\tilde{I}}
\newcommand{\tG}{\tilde{G}}
\newcommand{\tF}{\widetilde{F}}
\newcommand{\tN}{\tilde{N}}
\newcommand{\tU}{\tilde{U}}
\newcommand{\tV}{\tilde{V}}
\newcommand{\tK}{\tilde{K}}
\newcommand{\tY}{\tilde{Y}}
\newcommand{\tQ}{\tilde{Q}}
\newcommand{\tH}{\tilde{H}}
\newcommand{\tS}{\tilde{S}}

\newcommand{\tq}{\tilde{q}}
\newcommand{\tj}{\tilde{j}}
\newcommand{\tm}{\tilde{m}}
\newcommand{\tl}{\tilde{l}}
\newcommand{\tx}{\tilde{x}}
\newcommand{\ty}{\tilde{y}}
\newcommand{\tchi}{\tilde{\chi}}
\newcommand{\tell}{\tilde{\ell}}
\newcommand{\tal}{\tilde{\al}}
\newcommand{\tbeta}{\tilde{\beta}}
\newcommand{\ta}{\tilde{a}}
\newcommand{\tsigma}{\tilde{\sigma}}
\newcommand{\tep}{\tilde{\epsilon}}
\newcommand{\tla}{\tilde{\lambda}}
\newcommand{\tpsi}{\tilde{\psi}}

\newcommand{\tr}{\mbox{Tr}}

\newcommand{\NS}{\mbox{NS}}
\newcommand{\tNS}{\widetilde{\mbox{NS}}}
\newcommand{\R}{\mbox{R}}
\newcommand{\tR}{\widetilde{\mbox{R}}}
\newcommand{\sNS}{\msc{NS}}
\newcommand{\stNS}{\widetilde{\msc{NS}}}
\newcommand{\sR}{\msc{R}}
\newcommand{\stR}{\widetilde{\msc{R}}}

\newcommand{\any}{{}^{\forall}}
\newcommand{\ex}{{}^{\exists}}

\newcommand {\eqn}[1]{(\ref{#1})}

%%%%%%%%%%%%%%%%%%%%%%%%%%%%%%%%%%%%%

%%%%%%%%%%%%%%%%%%%%%%%%%%%%%%%%%%%%%%%%%%%%%%%%%%%%%%%%
%%%%%%%%%%%%%%%%%%%%%%%%%%%%%%%%%%%%%%%%%%%%%%%%%%%%%%%%
%%%%%%%%%%%%%%%%%%%%%%%%%%%%%%%%%%%%%%%%%%%%%%%%%%%%%%%%

%%% Title page %%%%%
\begin{titlepage}
 \
 \renewcommand{\thefootnote}{\fnsymbol{footnote}}
 \font\csc=cmcsc10 scaled\magstep1
 {\baselineskip=16pt
  \hfill
 \vbox{\hbox{May, 2016}
       }}

 \baselineskip=20pt
\vskip 1cm
 
\begin{center}

{\bf \Large

More on Non-supersymmetric Asymmetric Orbifolds

with 

Vanishing Cosmological Constant 

} 

 \vskip 1.2cm
 
 %\medskip

%\vskip 8mm
\noindent{ \large Yuji Sugawara}\footnote{\sf ysugawa@se.ritsumei.ac.jp},
%%%
\hspace{1cm}
\noindent{ \large Taiki Wada}\footnote{\sf rp0017xp@ed.ritsumei.ac.jp},

\medskip

 {\it Department of Physical Sciences, 
 College of Science and Engineering, \\ 
Ritsumeikan University,  
Shiga 525-8577, Japan}

\end{center}

\bigskip

~

\begin{abstract}

We explore various non-supersymmetric type II string vacua constructed based on
asymmetric orbifolds of tori with vanishing cosmological constant at the one loop. 
The string vacua we present are modifications of the models studied in \cite{SSW},
of which orbifold group is just generated  by a single element. 
We especially focus on two types of modifications: 
{\bf (i)} the orbifold twists include different types of 
chiral reflections not necessarily removing massless Rarita-Schwinger fields in the 
4-dimensional space-time,
{\bf (ii)} the orbifold twists do not include the shift operator.
We further discuss the unitarity and stability of constructed non-supersymmetric string vacua,
with emphasizing the common features of them.

\end{abstract}

\setcounter{footnote}{0}
\renewcommand{\thefootnote}{\arabic{footnote}}

\end{titlepage}

\baselineskip 18pt

\vskip2cm 
\newpage

%%%%%%%%%%%%%%%%%%%%%%%%%%%%%%%%%%%%%%%%
%%%%%%%%%%%%%%%%%%%%%%%%%%%%
%%%%%%%%%%%%%%%%%%%%%%%%%%%%%%%%%%%%%%%%
%%%%%%%%%%%%%%%%%%%%%%%%%%%%

%-----------------------------------------------
%%%   contents   %%%
%\setcounter{tocdepth}{3}
\tableofcontents

%%%%%%%%%%%%%%%%%%%%%%%%%%%%%%%%%%%%%%%%%%%%%%%%%%%%%%%%%%%%%%%%%%%%%%%%%%%%%
%%%%%%%%%%%%%%%%%%%%%%%%%%%%%%%%%%%%%%%%%%%%%%%%%%%%%%%%%%%%%%%%%%%%%%%%%%%%%

\section{Introduction and Summary}

Much attention has been currently focused on the string theories on non-geometric backgrounds. 
A simple and interesting class of such backgrounds are constructed due to the asymmetric orbifolds, 
in which the orbifold twists act asymmetrically on the left and right movers \cite{Narain:1986qm}. 
Although they look beyond our intuitive picture of space-time, they are 
well-described as done for geometrical ones
by the approach of world-sheet conformal field theory (CFT) in the $\al'$-exact fashion. 
%%%

Above all, one of the natural purposes to study the 
%compactifications of
type II string on asymmetric orbifolds  would be the construction of 
non-supersymmetric (SUSY) string vacua {\em with vanishing cosmological constant\/} motivated by 
phenomenological or theoretical interests.
It seems evident that the SUSY-breaking realized in any geometric or symmetric orbifolds 
inevitably gives rise to a non-vanishing cosmological constant already at the one-loop. 
In this sense, the bose-fermi cancellation without SUSY would only be 
possible in the suitable non-geometric compactification in superstring theory. 
The attempts of construction of non-SUSY vacua 
have been initiated by the works \cite{Kachru1,Kachru2,Kachru3} 
based on some non-abelian orbifolds, followed by closely related studies {\em e.g.} in 
\cite{Harvey,Shiu-Tye,Blumenhagen:1998uf,Angelantonj:1999gm}. 
%%%%%%%%%%%%%%%%%%
Moreover, sharing similar motivations, 
non-SUSY vacua in heterotic string theory have been investigated 
{\em e.g.} in
\cite{Blaszczyk:2014qoa,Angelantonj:2014dia,Faraggi:2014eoa,Abel:2015oxa,Angelantonj:2015nfa}.

%%%%%%%%%%%%%%%%%%%%%%%%%%%%%%%%%%%%%%%%%%%%%%
%%%%%%%%%%%%%%%%%%%%%%%%%%%%%%%%%%%%%%%%%%%%%%

Recently,  in our previous paper \cite{SSW},
we have presented a simple new realization of non-SUSY string vacua 
with the bose-fermi cancellation based on a {\em cyclic orbifold\/}, that is, 
the relevant orbifold group is generated by a single element.  
Hence, this construction  looks rather simpler than the previous ones given in the papers quoted above.  
The crucial point in this construction is the fact that  `chiral reflection' (or the T-duality twist) 
along the $T^4$-directions\footnote
   {Through this paper, $X^{\mu} \equiv (X^{\mu}_L, X^{\mu}_R)$ ($\psi^{\mu}\equiv (\psi^{\mu}_L, \psi^{\mu}_R)$) denotes 
the world-sheet bosonic (fermionic) fields in the RNS formalism of type II string theory. 
The directions $\mu=0,\ldots, 3$ are always identified as the 4-dim. Minkowski space-time $M^4$, and we mainly focus on the transverse part $\mu=2, \ldots, 9$.
In addition, we will often use the notations $\la^i\equiv (\la^i_L, \la^i_R)$ ($i=1,\ldots, 2N$) to express the free fermions describing 
the $N$-dim. torus with the $SO(2N)$-symmetry enhancement, which will be denoted as $T^N[SO(2N)]$ in the text.  
};
\begin{align}
\cR \equiv (\-_R)^{\otimes 4}
~ : ~ 
&(X_L^{\mu}, X_R^{\mu})~ \longmapsto ~ (X_L^{\mu}, -X_R^{\mu}), 
%(\mu=6,7,8,9),
\nonumber\\
&(\psi_L^{\mu}, \psi_R^{\mu})~ \longmapsto ~ (\psi_L^{\mu}, -\psi_R^{\mu}), 
%\\
%& 
\hspace{1cm} (\mu=6,7,8,9),
\label{def chiral reflection}
\end{align}
is {\em not\/} necessarily involutive when acting on the world-sheet fermions, even in the untwisted sector\footnote{
   It is well-known that the chiral reflections often define order $N \geq 4$ orbifolds rather than order 2
due to the non-trivial phase factors appearing in the {\em twisted sectors}, even though they act as an involution on 
the untwisted sector. 
See {\em e.g.} \cite{Aoki:2004sm}. 
}.
Indeed, as illustrated in \cite{SSW}, 
while it is always involutive on the (right-moving) NS-fermions in the untwisted sector, 
we still have two possibilities 
{\bf (i)} $\cR^2 = {\bf 1}$, {\bf (ii)} $\cR^2 = \-$
for the R-sector.
%%%%%%%%
%%%%%%%%
In other words, even though
$\cR^2$ obviously commutes with all the world-sheet coordinates; 
$$\cR^2 X^{\mu}_R \cR^{-2} = X^{\mu}_R,
\hspace{1cm} 
\cR^2 \psi^{\mu}_R \cR^{-2} = \psi^{\mu}_R,
$$
it may still act on the Ramond vacua (or spin fields) as a  sign flip.
%%%%%%%
%%%%%%%
%In other words, 
The case {\bf (ii)} means that $\cR^2 = (-1)^{F_R}$,
where $F_R$ ($F_L$) denotes the `space-time fermion number' from the right(left)-mover.
%Throughout this paper, we shall adopt the second one.
If taking the second one, which we often call the `$\bz_4$-chiral reflection',
one  finds that the type II string vacuum constructed as the  $\bz_4$-orbifold by  
%\begin{equation}
$
\sigma \equiv (-1)^{F_L} \otimes (\-_R)^{\otimes 4}
$
%\end{equation}
possesses the next properties;
\begin{itemize}
\item All the space-time supercharges arising from the untwisted sector are eliminated by the $\bz_4$-projection
$
\frac{1}{4} \sum_{r\in \bz_4}\, \sigma^r,
$
since any supercharges in the unorbifolded theory do not commute with both of $(-1)^{F_L}$ and $(-1)^{F_R}$.

\item
All the partition sums in the untwisted sector vanish under the insertion of $\sigma^r$ for  $\any r \in \bz_4$.
Namely, we find
%%%%
($q \equiv e^{2\pi i \tau}$);
%%%%
$$
\tr_{\msc{untwisted}} \left[\sigma q^{L_0-\frac{c}{24}} \bar{q}^{\tL_0-\frac{c}{24}}\right] = 
\tr_{\msc{untwisted}} \left[\sigma^3 q^{L_0-\frac{c}{24}} \bar{q}^{\tL_0-\frac{c}{24}}\right]=0,
$$
due to the cancellation in the right moving fermions caused by $(\-_R)^{\otimes 4}$, while 
$$
\tr_{\msc{untwisted}} \left[q^{L_0-\frac{c}{24}} \bar{q}^{\tL_0-\frac{c}{24}}\right] = 
\tr_{\msc{untwisted}} \left[\sigma^2 q^{L_0-\frac{c}{24}} \bar{q}^{\tL_0-\frac{c}{24}}\right]=0,
$$
holds because $\sigma^2$ trivially acts on the left-mover, yielding the familiar vanishing factor 
$\theta^4_3- \theta^4_4- \theta^4_2 $. 
\end{itemize}
%%%%%
They are surely nice features for the purpose to realize the non-SUSY string vacua 
with the  bose-fermi cancellation. 
%%%%
However, as addressed in \cite{SSW} and will be demonstrated in section 2 for a  detail, it turns out that 
8 supercharges eventually emerge {\em in the twisted sector}.
%  in this $\bz_4$-orbifold.
We thus adopted in \cite{SSW} the (infinite order) orbifold group generated by the operator
\begin{equation}
g= \cT_{2\pi R} \otimes \sigma \equiv \cT_{2\pi R} \otimes (-1)^{F_L} \otimes (\-_R)^{\otimes 4} , 
\label{g previous}
\end{equation}
in place of $\sigma$, following the spirit of Scherk-Schwarz type compactification \cite{SS1,SS2}. 
Here, $\cT_{2\pi R}$ denotes the shift by $2 \pi R$ along the `base' direction, 
originally identified as a real line $\br_{\msc{base}}$.
%%%%
The inclusion of shift into  \eqn{g previous} enables us to naturally identify the twisted sectors with 
the winding sectors of the `Scherk-Schwarz circle'. 
More significantly, it plays the role of removing potential supercharges which might arise 
from the twisted sectors\footnote
   {At first glance, this fact would look obvious, since the inclusion of shift $\cT_{2\pi R}$ generically 
makes all the Ramond states lying in the twisted sectors massive. However, we often find that 
additional Ramond massless states appear 
when choosing the Scherk-Schwarz radius $R$ suitably. Nevertheless, one can show that the space-time SUSY is completely broken 
for an arbitrary value of $R$. See \cite{SSW} for the detail. 
}.
%%%%%%%%%%%%%%%%%%%%%%%%%%%%%%%%%%%%%%%%%%%%%%%%%%%%%%%%%%%%%%%%%%%%%%%%%%%
%%%%%%%%%%%%%%%%%%%%%%%%%%%%%%%%%%%%%%%%%%%%%%%%%%%%%%%%%%%%%%%%%%%%%%%%%%%
We also note that this model would be interpreted as a modification of
the simple realizations of the `T-folds' 
\cite{Dabholkar:2002sy,Hellerman:2002ax,Flournoy:2004vn,Hull:2004in,Shelton:2005cf,Shelton:2006fd,Dabholkar:2005ve}, 
that is, the orbifolds  by the chiral reflection (or the T-duality twist) 
combined with the shift in the base space. 
These types of non-geometric backgrounds  have been 
studied by the approach of  world-sheet CFT  {\em e.g.\/}  in
 \cite{Flournoy:2005xe,HW,KawaiS1,KawaiS2,
 Condeescu:2012sp,Condeescu:2013yma,SatohS,Tan:2015nja}.%
%

%%%%%%%%%%%%%%%%%%%%%%%%%%%%%%%%%%%%%%%%%%%%%%%%%%%%%%%%%%%%%%%%%%%%%%%%%%%
%%%%%%%%%%%%%%%%%%%%%%%%%%%%%%%%%%%%%%%%%%%%%%%%%%%%%%%%%%%%%%%%%%%%%%%%%%%
%%%%%%%%%%%%%%%%%%%%%%%%%%%%%%%%%%%%%%%%%%%%%%%%%%%%%%%%%%%%%%%%%%%%%%%%%%%

Now, in this paper,  we would like to explore a variety of non-SUSY string vacua of this type. 
We shall especially focus on the next two modifications of \eqn{g previous}:
%%%%%
\begin{description}
\item[(i)]
We replace $(-1)^{F_L}$ 
%in \eqn{g previous} 
with 
$(\-_L)^{\otimes 2}$, 
which acts along the various directions of back-gorund tori, and 
plays the role of breaking the left-moving SUSY.

\item[(ii)]
We do not include the shift operator $\cT_{2\pi R}$. 
%in \eqn{g previous}.
Instead, we assume that $\cR \equiv (\-_R)^{\otimes 4}$ acts as the $\bz_4$-chiral reflection 
also for the world-sheet {\em bosons\/}. This is achieved by utilizing the fermionization of bosonic coordinates $X^{\mu}$,
and plays the role of preventing 
%new supercharges from arising from the twisted sectors. 
the twisted sectors from providing additional supercharges.

\end{description}

%%%%%%%%%%%%%%%%%%%%%%%%%%%%%%%%%%%%%%%%%%%%%%%%%%%%%%%%%%%%%%5

Stated more concretely, the models that we shall study in this paper 
are displayed in Tables \ref{7twists} and \ref{7backgrounds}.
In section 2, we briefly review on the `previous' one studied in \cite{SSW}, 
which would be helpful to readers. 
We then investigate the new six models (`models I to VI') in section 3. 
%%%%
We exhibit the relevant orbifold actions in Table \ref{7twists}, while 
the original backgrounds that we orbifold are summarized in Table \ref{7backgrounds}.
In all the models the orbifold groups are generated by a single element 
denoted as $g$ in Table \ref{7twists}.
%%%%%%
In Table \ref{7backgrounds}, 
$M^4$ expresses the four-dimensional Minkowski space-time.
The orbifold twists do not act on 
$\left[ M^4  \times \cdots\right]$ in each row.
The shift $\cT_{2\pi R}$ always acts along $\br_{\msc{base}}$. 
Throughout this paper, we use the notation `$T^N[SO(2N)]$' to express the $N$-dimensional torus at  
the symmetry enhancement point of $SO(2N)$. In other words, they can be described 
in terms of $2N$ Majorana fermions (denoted as `$\la^i\equiv (\la^i_L, \la^i_R)$').

%%%%%%%%%%%%%%%%%%%%%%%%%%%%%%%%%%%%%%%%%%%%%%%%%%%%%%%%%%%%%%%%%%%%%%
%%%%%%%%%%%%%%%%%%%%%%%%%%%%%%%%%%%%%%%%%%%%%%%%%%%%%%%%%%%%%%%%%%%%%%

%--------------------------------------------------------
\begin{table}[h]

\begin{center}
\caption{The orbifold actions}
\label{7twists}
\vspace{2mm}
{\renewcommand\arraystretch{1.4}
\begin{tabular}{crr}
\hline
model & \multicolumn{1}{c}{ $g$} & \multicolumn{1}{c}{$g^2$}\\
\hline
%---------previous--------
previous & $
\mathcal T_{2\pi R}
\otimes
 (-1)^{F_L}|_\psi
\otimes (-\mathbf 1_R)^{\otimes 4} 
$
&
$
\mathcal T_{4\pi R}
\otimes (-1)^{F_R}|_{\psi}
$\\
%---------I, II--------
I, II & $
\mathcal T_{2\pi R}
\otimes
 (-\mathbf 1_L)^{\otimes 2}
\otimes (-\mathbf 1_R)^{\otimes 4} 
$
&
$
\mathcal T_{4\pi R}
\otimes (-1)^{F_R}|_{\psi}
$\\
%---------III--------
III & $

 (-1)^{F_L}|_\psi
\otimes (-\mathbf 1_R)^{\otimes 4} 
$
&
$
(-1)^{F_R}|_{\lambda}
\otimes (-1)^{F_R}|_{\psi}
$\\
%---------IV,V,VI--------
IV, V, VI & $

 (-\mathbf 1_L)^{\otimes 2}
\otimes (-\mathbf 1_R)^{\otimes 4} 
$
&
$
(-1)^{F_R}|_{\lambda}
\otimes (-1)^{F_R}|_{\psi}
$\\
\hline
\end{tabular}}
\end{center}
\end{table}

%--------------------------------------------------------
\begin{table}[h]
%\tablinesep =5pt
\begin{center}
\caption{The original backgrounds.   }
\label{7backgrounds}
\vspace{2mm}
{\renewcommand\arraystretch{1.4}
\begin{tabular}{cl}
\hline
models & original backgrounds \\
\hline
%---------previous--------
previous
&
 $\left[ M^{4}\times S^1 \right] \times \mathbb R
_{\mathrm{base}} 
\times T^4
[SO(8)]
  $
\\
\hdashline
%---------model I--------
I& 
 $\left[ M^{4}\times S^1 \right] \times \mathbb R
_{\mathrm{base}} 
\times T^2\times T^2
[SO(4)]
  $
\\
%\hline

%---------model II--------
II
&
 $\left[ M^{4} \right]\times\mathbb R
_{\mathrm{base}} 
\times  S^1 \times T^4
[SO(8)]
  $
\\
\hdashline
%--------model III--------
III& 
 $\left[ M^{4}\times T^2 \right] 
\times T^4
[SO(8)]
  $
\\
\hdashline
%--------model IV--------
IV&
 $\left[ M^{4}\times T^2 \right] \times T^2
\times T^2
[SO(4)]
$
\\
%\hline
%--------model V--------
V& 
 $\left[ M^{4}\times S^1 \right] \times S^1
\times T^4
[SO(8)]
  $
\\
%\hline
%--------model VII--------
VI
&
 $\left[ M^{4} \right]
\times T^6
[SO(12)]
$
\\
\hline

\end{tabular}
}
\end{center}
\end{table}

%%%%%%%%%%%%%%%%%%%%%%%%%%%%%%%%%%%%%%%%%%%%%%%%%%%%%
%%%%%%%%%%%%%%%%%%%%%%%%%%%%%%%%%%%%%%%%%%%%%%%%%%%%%

Let us summarize the aspects of models I to VI on which 
we will elaborate in section 3. 
%%%%
The models I and II are defined by including $(\-_L)^{\otimes 2}$ instead of $(-1)^{F_L}|_\psi$.
Combining it with $(\-_R)^{\otimes 4}$, some directions of tori are eventually orbifolded by the non-chiral reflection:
$ (X^{\mu}_L , X^{\mu}_R) \, \rightarrow \, (-X^{\mu}_L, - X^{\mu}_R)$, and 
we simply denote `$T^2$' and `$S^1$' for the corresponding directions. 
%on which the non-chiral reflections act, since we need not specify the moduli for these directions. 
It will be shown that these models are indeed the non-SUSY string vacua with the bose-fermi cancellation as expected.
We do not have any tachyonic instability in all the untwisted and twisted sectors, while 
some winding massless modes emerge at particular values of the Scherk-Schwarz radius $R$. 
These features are quite similar to the previous one.
%%%
However, the physical spectra significantly differ from it. 
Some Rarita-Schwinger fields survive in the 4-dim. massless spectrum in the models I and II,
although not interpreted as the gravitini due to the absence of space-time SUSY. 
We recall that, in the previous model, the twist  by $(-1)^{F_L}$ eliminates
all the massless spin 3/2 states in the untwisted sector. 
%as shown in \cite{SSW}.

%%%%%%%%%%%%%%%%%%%%%%%

The models III-VI  are those not including  the shift operator. 
Instead, we shall modify the right-moving chiral reflections 
%$(\-_R)^{\otimes 4}$ or $(\-_R)^{\otimes 2}$ 
 so that their squares yield $(-1)^{F_R}|_{\lambda }$, that is, 
the sign flip on the Ramond sector of  
fermions  $\la_R^i$ that describe  $T^N[SO(2N)]$.
%%%%%%
The left-moving space-time SUSY is broken by $(-1)^{F_L}|_{\psi}$ 
in the model III as in the previous one, while 
$(\-_L)^{\otimes 2}$ acts on the various directions of tori in the cases of models IV-VI.
%%%
By the effect of $(-1)^{F_R}|_{\lambda }$, 
the twisted sectors gain extra zero point energies despite the absence of shift operator,  
thereby preventing additional right-moving supercharges from arising.
It then turns out that we achieve the desired non-SUSY vacua. 
They are simpler than the models I and II for the computations of the torus partition functions.
Once again, we do not face any tachyonic instabilities, and massless states appear
in the twisted sectors as well as the untwisted sector.
%%%
Note that these models do not include the modulus $R$ as opposed to the cases of models I and II.  
%%%
%%%%%%%%%%%%%%%%%%%%%%%%%%%%%%%%%%%

The partition functions for all the models in this paper are found 
%to be 
manifestly  modular invariant  and  
$q$-expanded in the way compatible with unitarity. 
Moreover, they are always free from tachyonic instabilities.  
These would be common features of the toroidal asymmetric orbifolds of these types,
as we will discuss  in section \ref{Discussion}.

%%%%%%%%%%%%%%%%%%%%%%%%%%%%%%%%%

~

%%%%%%%%%%%%%%%%%%%%%%%%%%%%%%%%%%%%%%%%%%%%%%%%%%%%%%%%%%%%%%%%%%%%%%%%%%%%%%%%%%%%%%%%%%%
%%%%%%%%%%%%%%%%%%%%%%%%%%%%%%%%%%%%%%%%%%%%%%%%%%%%%%%%%%%%%%%%%%%%%%%%%%%%%%%%%%%%%%%%%%%
%%%%%%%%%%%%%%%%%%%%%%%%%%%%%%%%%%%%%%%%%%%%%%%%%%%%%%%%%%%%%%%%%%%%%%%%%%%%%%%%%%%%%%%%%%%
%-----------------------------------------------------------------------------------
\section{Notes on the Non-SUSY Asymmetric Orbifold of \cite{SSW}}
\label{Construction}

In this section, we make a brief sketch of the non-SUSY model constructed in \cite{SSW} to clarify several points
that we will discuss for  the new models.

Let us introduce the type II string vacuum in the ten-dimensional flat background;
\begin{equation}
\left[M^{4}\times S^1\right] \times \mathbb R
_{\mathrm{base}}
\times T^4
%_{\mathrm{fiber}}
[SO(8)] , \label{bg0}
\end{equation}
where $M^{4}$ ($X^{0,1,2,3}$ -directions) denotes the 4-dimensional Minkowski space-time, and
$S^1$ ($X^4$ -direction) is a circle that plays no role in this model.
$\mathbb R 
_{\mathrm{base }}
$
($X^5$ -direction)
 is just a real line, identified as the `base space' of the twisted compactification like 
Scherk-Schwarz \cite{SS1,SS2},
and, as already mentioned,
$T^4
% _{\mathrm{fiber}}
[SO(8)]$ ( $X^{6,7,8,9}$ -directions) is 
 the 4-dimensional torus with the $SO(8)$-symmetry
enhancement.

Then, as was introduced in section 1, 
we define the asymmetric orbifold generated by the operator
\begin{equation}
{g} = 
\mathcal T_{2\pi R}
%|_{
%X^5
%\mathbb R_\mathrm{base}
%} 
\otimes
\sigma 
\equiv
\mathcal T_{2\pi R}
%|_{
%%X^5
%\mathbb R_\mathrm{base}
%}  
\otimes (-1)^{F_L}|_\psi 
\otimes (-\mathbf {1}_R)^{\otimes 4}|_{T^4},
\label{asym action0}
\end{equation}
acting on the background \eqref{bg0}.
Recall that 
%%%%%
$\cT_{2\pi R} $ denotes the shift operator along $\br_{\msc{base}}$; $X^5\, \rightarrow X^5 + 2\pi R$,
%%%%%
and 
the operator $(-1)^{F_L}|_{\psi}$ ($(-1)^{F_R}|_{\psi}$) acts as the sign flip of the left (right) moving Ramond sector.
$(-\mathbf {1}_R)^{\otimes 4}|_{T^4}$ denotes the chiral reflection along $T^4$ given in \eqn{def chiral reflection}.
%%%%%%%%%
%%%%%%%%%
To complete the definition of the operator $\sigma$ (or $(-\mathbf {1}_R)^{\otimes 4}|_{T^4}$), 
we still need to specify the construction of Ramond vacua (or spin fields) of right-moving world-sheet fermions $\psi^{\mu}_R$ and 
how $\sigma$ should act on them. 
Here, we define the Ramond vacua as 
$\ket{s_1, \ldots, s_4}_{R} \equiv e^{i\sum_{a=1}^4 s_a H^a_{R}}\ket{0}_R $,  
($s_a \equiv \pm \frac{1}{2}$),  where  $e^{i\sum_{a=1}^4 s_a H^a_{R}}$ 
denotes the $SO(8)$-spin fields associated to the transverse fermions  $\psi_R^2, \ldots, \psi_R^9$
by the bosonization
\begin{align}
& \psi_R^2\pm i\psi_R^3 = \sqrt{2} e^{\pm i H_R^1}, ~~~ \psi_R^4\pm i\psi_R^6 = \sqrt{2} e^{\pm i H_R^2}, 
\nonumber
\\
&
\psi_R^5\pm i\psi_R^7 = \sqrt{2} e^{\pm i H_R^3}, ~~~ \psi_R^8\pm i\psi_R^9 = \sqrt{2} e^{\pm i H_R^4}.
\label{bosonization}
\end{align}
We then obtain 
\begin{align}
\sigma \ket{s_1,s_2,s_3,s_4}_R = 
e^{ i\pi s_4}   \ket{s_1,-s_2,-s_3,s_4}_R,
\label{def sigma right f}
\end{align}
since $\sigma$ acts as the sign flip of $\psi_R^6, \ldots \psi_R^9$.
%%%%%%%%%%
Thus we readily find
\begin{equation}
\sigma^2 = (-1)^{F_R}|_\mathrm{\psi},
\end{equation}
which plays a crucial role in the following discussions. See \cite{SSW} for more detail.
%%%%%%%%%%%%%%%%%%%%%

Let us  focus on the partition function on the world-sheet torus to investigate
the one-loop cosmological constant and the space-time supersymmetry. 
The relevant partition function is schematically written in the form as 
\begin{equation}
Z(\tau , \bar \tau )=\sum_{w,m \in \mathbb Z}
{Z}_{(w,m)}(\tau,\bar \tau )\equiv \sum_{w,m \in \mathbb Z}
{Z}^{X}_{(w,m)}(\tau,\bar \tau )
Z^{\psi_L} _{(w,m)}(\tau) 
\overline{Z^{\psi_R} _{(w,m)}(\tau)} ,
%Z^{\tilde{\psi}} _{(w,m)}(\bar \tau) 
\label{sample partition}
\end{equation}
%%%%
%%%%
where the integer $w$ labels the twisted sectors, while $m$ indicates the $g^m$-insertions into the trace.
As already suggested in section 1, they are identified as the spatial and temporal winding numbers on
the base space (or the Scherk-Schwarz circle) because of 
the inclusion of shift $\cT_{2\pi R}$ into \eqn{asym action0}.
%%%%
%%%%
${Z}^{X}_{(w,m)}(\tau,\bar \tau )$ denotes the partition functions of the bosonic sectors,
while $Z^{\psi_L} _{(w,m)}(\tau)$, $Z^{\psi_R} _{(w,m)}(\tau)$ are the partition functions 
of the left- and right-moving fermionic sectors. 
%%%

Each partition sum ${Z}_{(w,m)}(\tau,\bar \tau )$ is evaluated in the easiest way  as follows.
We first calculate the trace over the untwisted sector $(w=0)$\footnote
   {Here we shall adopt the conventional normalization 
of the trace for the CFT describing $\br_{\msc{base}}$; 
$$
\tr \left[q^{L_0-\frac{1}{24}} \overline{q^{\tL_0-\frac{1}{24}}}\, \right]
=\frac{R}{\sqrt{\tau_2} \left|\eta\right|^2},
$$
so that we simply obtain
$$
\tr \left[\left(\cT_{2\pi R}\right)^m \,q^{L_0-\frac{1}{24}} \overline{q^{\tL_0-\frac{1}{24}}}\, \right]
=\frac{R}{\sqrt{\tau_2} \left|\eta\right|^2} e^{- \frac{\pi}{\tau_2} R^2 m^2}.
$$
%This means that we start with $S^1_{NR}$ for the base CFT
%with an arbitrary integer $N$, and 
%regard the insertion of the shift operators $(\cT_{2\pi R})^m$ as 
% implementing 
%the $\bz_N$-orbifolding. 
} , 
%($q \equiv e^{2\pi i \tau}$) 
\begin{align}
Z_{(0, m)}(\tau , \bar \tau) & =\mathrm{Tr}_{w=0}
\left[ {g} ^m q^{L_0 -\frac{c}{24}} \bar q^{\tilde L_0 -\frac{c}{24}} \right]
\nonumber \\
& =  
Z_{R, (0,m)} (\tau , \bar \tau) \,
\mathrm{Tr}_{w=0}\,
\left[ {\sigma} ^m q^{L_0 -\frac{c}{24}} \bar q^{\tilde L_0 -\frac{c}{24}} \right],
%\ \ \ q\equiv e^{2\pi i \tau}
\\
Z_{R, (w,m)} (\tau , \bar \tau) 
& \equiv \frac{R}{\sqrt \tau _2 |\eta (\tau)|^2}e^{-\frac{\pi R^2}{\tau _2}|w\tau +m|^2}, \ \ \ 
(w,m \in \mathbb Z), 
\label{Rblock}
\end{align}
and those for the general winding sectors $(w,m)$ are uniquely determined by 
requiring the modular covariance
\begin{align}
&Z_{(w,m)}(\tau , \bar \tau)|_S
=Z_{(m,-w)}(\tau , \bar \tau ) , \\
&Z_{(w,m)}(\tau , \bar \tau)|_T
=Z_{(w,w+m)}(\tau , \bar \tau ) ,
\end{align} 
where  $S: \tau \to -1/\tau $, $T: \tau \to \tau +1 $ are the modular transformations.
We then achieve the partition function \eqref{sample partition} that is manifestly modular invariant.

%%%%%%%%%%%%%%%%%%%%%%%%%%%%%%%%%%%%%%%%%%%%%%%%%%%%%%%%%%%%%%

Note that the left and right partition sums of fermionic sectors  $Z_{(w,m)}^{\psi_L}(\tau)$, 
$\overline{Z_{(w,m)}^{\psi_R}(\tau)}$ are generically asymmetric. 
%%%
%For the model defined by \eqref{asym action0}, 
The twist operator 
$\sigma $ includes $(-1)^{F_L} |_{\psi}$, and we thus find 
$ %\begin{align}
Z^{\psi_L} _{(w,m)}(\tau) \neq 0$ for $ ^\forall w \in 2\mathbb Z+1
$ or $ ^\forall m \in 2\mathbb Z+1
$. 
Similarly, by $\sigma ^2 = (-1)^{F_R}|_{\psi}$, we obtain
$ %\begin{align}
\overline{Z^{\psi_R} _{(w,m)}(\tau)} \neq 0$ for $ ^\forall w \in 4\mathbb Z+2
$ or $ ^\forall m \in 4\mathbb Z+2
$. 
However, one easily finds 
\begin{align}
\begin{array}{cl}
\overline{Z^{\psi_R} _{(w,m)}(\tau)} = 0,  & ( w\ 
 \mathrm{or}\ m \in 2\mathbb Z+1 ),\\
Z^{\psi_L} _{(w,m)}(\tau) = 0, &( w,m \in 2\mathbb Z) .
\end{array}
\end{align}
Thus the total partition function vanishes.

%%%%%%%%%%%%%%%%%%%%%%%%%%%%%%%%%%%%%%%%%%%%%%%%%%%%%%%%%%%%%

Let us turn our attention to the spectrum in the untwisted sector $(w=0)$. 
%%%%
As already mentioned in section 1, 
all the space-time supercharges are eliminated by 
the orbifold projection $\frac{1}{4}\sum _{n\in \mathbb Z_4} \sigma ^n$
due to the inclusions  $(-1)^{F_L}|_{\psi}$ and $(-1)^{F_R}|_{\psi}$.
For all that, one can observe that the same number of bosonic and fermionic states exist at each mass level of the untwisted sector.
%%%
Especially, the massless spectrum is summarized in 
Table \ref{mlspect1}, which includes 32 bosonic and fermionic states.
Note that no gravitino appears in the 4-dim. spectrum, 
which suggests the absence of space-time SUSY.

%%%%%%%%%%%%%%%%%%%%%%%%%%%%%%%%%%%%%%%%%%%%%%%%%%%%%

\begin{table}[h]
\begin{center}
	\caption{Massless spectrum in the untwisted sector for the orbifold model defined by 
%\eqref{asym action0}
${g} $}
\label{mlspect1}
{\renewcommand\arraystretch{1.25}
\vspace{2mm}
\begin{tabular}{|c|c|}
\hline
spin structure & 4D fields\\
\hline
\hline
 (NS, NS) &  graviton, 8 vectors,\\
 &  14 (pseudo) scalars\\
\hline 
 (R , NS)  & 16 Weyl fermions\\
\hline
\end{tabular}}
\end{center}
\end{table}

%%%%%%%%%%%%%%%%%%%%%%%%%%%%%%%%%%%%%%%%%%%%%%%%%%%%%%%%%%

However, this is not the whole story. 
It might be possible that new supercharges arise from the twisted sectors.
We also note that tachyonic states would potentially  emerge in the twisted sectors, as in 
many examples of the SUSY-breaking models of Scherk-Schwarz type. 
Furthermore, the unitarity of string spectrum is not necessarily self-evident 
because of the non-trivial phase factors appearing in the twisted sectors necessary for the modular invariance. 
It is surely significant to examine these issues for our purpose. 
A direct way to do so is to decompose the partition functions with respect to the spatial
winding $w$ and the spin structures as 
\begin{align}
& Z(\tau , \bar \tau )= \frac{1}{4} \cZ_{M^4 \times S^1} (\tau,\bar{\tau}) 
\nonumber \\
& \hspace{1cm} \times 
\sum _{w \in \mathbb Z} \left \{
 Z_w^{(\mathrm{NS}, \mathrm{NS})}(\tau , \bar \tau )
+Z_w^{(\mathrm{NS}, \mathrm{R})}(\tau , \bar \tau )
+Z_w^{(\mathrm{R}, \mathrm{NS})}(\tau , \bar \tau )
+Z_w^{(\mathrm{R}, \mathrm{R})}
(\tau , \bar \tau )\right \},
\label{departition0}
\end{align} 
where $\cZ_{M^4 \times S^1}$ denotes the bosonic transverse contribution for the $M^4 \times S^1$-sector 
that has nothing to do with the orbifolding. 
%%%
The string spectrum in each Hilbert space with winding $w$
can be examined by making the Poisson resummation
with respect to the temporal winding $m$.
%%%
In this way 
the following results have been shown in \cite{SSW}; 
%%%%%%%%%%%%%%%%%%%%%%%%%%%%%%%%%%%%%%%%%%%%%%%%%%%%%%%%%
\begin{itemize}
 \item
%[$\Box$] 
The partition function for each winding $w$ and each spin structure is compatible with unitarity. 
 \item
%[$\Box$] 
The bose-fermi cancellation is observed at each mass level of the string spectrum.
%%%%
\item 
The space-time SUSY is completely broken.
%%%
 \item
%[$\Box$] 
No tachyonic states appear in all the  sectors.
 \item
%[$\Box$] 
Massless states arise in some twisted sectors
at the specific radius $R$ (the modulus related to the shift $\cT_{2\pi R}$). 
\end{itemize}
%%%%%%%%%%%%%%%%%%%%%%%%%%%%%%%%%%%%%%%%%%%%%%%%%%%%%%%%%

Especially, let us focus on  how one can conclude that the space-time SUSY is truly broken.
It has been explicitly shown in \cite{SSW} that the partition functions for the winding sectors have 
the relations summarized in Table \ref{all partition relation}. 
For the odd winding sectors,
%($w\in 2\bz+1$), 
we have the bose-fermi cancellation compatible only with right-moving SUSY,
while the even sectors 
%($w\in 2\bz$)  
behave as if we only had left-moving supercharges.
It is obvious that any supercharges  can never be consistent
with both of them at the same time.
%the relations depicted in table \ref{all partition relation}.
%We have thus achieved a non-SUSY vacuum  with the bose-fermi cancellation.
%%%%%%%%%%%%%%%%%%%%%%%%%%%%%%%
\begin{table}[ht]
\begin{center}
	\caption{Relations among  the winding sectors in the orbifold 
defined by \eqn{asym action0}. ($\any w' \in \mathbb Z$) }
\label{all partition relation}
\vspace{2mm}
	\begin{tabular}{ll|l}
%\hline
%&left & right \\
%\hline
%\hline
$w\in 2\mathbb Z +1 $ &&\\
& $Z^{(\mathrm{NS}, \mathrm{NS})}_{w} \neq -Z^{(\mathrm{R}, \mathrm{NS})}_{w'} $
& $Z^{(\mathrm{NS}, \mathrm{NS})}_{w} = -Z^{(\mathrm{NS}, \mathrm{R})}_{w} $\\
&$Z^{(\mathrm{NS}, \mathrm{R})}_{w} \neq -Z^{(\mathrm{R}, \mathrm{R})}_{w'} $ 
& $Z^{(\mathrm{R}, \mathrm{NS})}_{w} = -Z^{(\mathrm{R}, \mathrm{R})}_{w} $\\
\hline 
$ w\in 2\mathbb Z $ &&\\
& $Z^{(\mathrm{NS}, \mathrm{NS})}_{w} = -Z^{(\mathrm{R}, \mathrm{NS})}_{w} $ 
& $Z^{(\mathrm{NS}, \mathrm{NS})}_{w} \neq -Z^{(\mathrm{NS}, \mathrm{R})}_{w'} $ \\
& $Z^{(\mathrm{NS}, \mathrm{R})}_{w} = -Z^{(\mathrm{R}, \mathrm{R})}_{w} $
& $Z^{(\mathrm{R}, \mathrm{NS})}_{w} \neq -Z^{(\mathrm{R}, \mathrm{R})}_{w'} $\\
\end{tabular}
\end{center}
\end{table}

%%%%%%%%%%%%%%%%%%%%%%%%%%%%%%%%%%%%%%%%%%%%%%%%%%%%%%%%%%%%%%%%%%%%%
%%%%%%%%%%%%%%%%%%%%%%%%%%%%%%%%%%%%%%%%%%%%%%%%%%%%%%%%%%%%%%%%%%%%%

~

%%%%%%%%%%%%%%%%%%%%%%%%%%%%%%%%%%%%%%%%%%%%%%%%%%%%%%
%%%%%%%%%%%%%%%%%%%%%%%%%%%%%%%%%%%%%%%%%%%%%%%%%%%%%%
\subsection*{Remarks on the supersymmetric cases}

It would be worthwhile to figure out what happens in the closely related model with the SUSY unbroken, that is, 
the asymmetric orbifold defined by 
$\sigma \equiv (-1)^{F_L}|_{\psi} \otimes (\-_R)^{\otimes 4}$ without including the shift. 
%We also assume 
%$\sigma ^2 = (-1)^{F_R}|_{\psi} $, 
%%%%
%%%%
We also adopt \eqn{def sigma right f} for the action of $\sigma$ on Ramond vacua,
%%%%
%%%%
and thus the orbifold twist is still a $\mathbb Z_4$-action. 
The partition function is then  written in the form as 
\begin{equation}
Z(\tau , \bar \tau )=\frac{1}{4}\sum_{a,b  \in \mathbb Z_4}
{Z}_{(a,b)}(\tau,\bar \tau )\equiv \frac{1}{4} \sum_{a,b  \in \mathbb Z_4}
{Z}^{X}_{(a,b)}(\tau,\bar \tau )
 Z^{\psi_L} _{(a,b)}(\tau) \overline{Z^{\psi_R} _{(a,b)}(\tau)}. 
\label{sample partition'}
\end{equation}
In this case, the orbifold projection 
still removes all the supercharges  in the untwisted sector, 
but the right-moving supercharges revive from the $a=2$ twisted sector.

To show this fact explicitly, let us again  decompose the partition functions as 
\begin{align}
& Z(\tau , \bar \tau )= \frac{1}{16} \cZ_{M^4 \times T^2} (\tau,\bar{\tau})
\nonumber \\
& \hspace{1cm} 
\times
\sum _{a \in \mathbb Z_4} \left \{
 Z_a^{(\mathrm{NS}, \mathrm{NS})}(\tau , \bar \tau )
+Z_a^{(\mathrm{NS}, \mathrm{R})}(\tau , \bar \tau )
+Z_a^{(\mathrm{R}, \mathrm{NS})}(\tau , \bar \tau )
+Z_a^{(\mathrm{R}, \mathrm{R})}(\tau , \bar \tau )
\right \},
\label{departitionO'}
\end{align}
where the overall factor $\frac{1}{16} \equiv \frac{1}{4} \times \frac{1}{4}$ 
is due to the $\bz_4$-orbifolding as well as the chiral GSO projection.
%%%
Then we obtain
%the following relations of the partition functions among all the sectors:
\begin{align}
&Z_0^{(\mathrm{NS} ,\mathrm{NS})}
(\tau , \bar \tau ) = -Z_0^{(\mathrm{R} ,\mathrm{NS})}
(\tau , \bar \tau )
=Z_2^{(\mathrm{R} ,\mathrm{R})}
(\tau , \bar \tau ) = -Z_2^{(\mathrm{NS} ,\mathrm{R})}
(\tau , \bar \tau)
 \nonumber \\
& ~~~ = Z_{1,3}^{(\mathrm{R} ,\mathrm{R})}
(\tau , \bar \tau ) = -Z_{1,3}^{(\mathrm{R} ,\mathrm{NS})}
(\tau , \bar \tau )
=\left \{ \left|\frac{\theta _3}{\eta}\right |^8
+ \left|\frac{\theta _4}{\eta}\right|^8  + \left |\frac{\theta _2}{\eta} \right |^8 \right \}
\left |\frac{\theta _2}{\eta} \right |^8 
\label{partitions without shift1} ,
\\
&Z_{1,3}^{(\mathrm{NS} ,\mathrm{NS})}
(\tau , \bar \tau ) = -Z_{1,3}^{(\mathrm{NS} ,\mathrm{R})}
(\tau , \bar \tau )
\nonumber \\
& ~~~ =
\overline{\left( \frac{\theta _2}{\eta } \right)^4} 
\left\{ 
\left| \frac{\theta _3}{\eta } \right|^8
-\left| \frac{\theta _4}{\eta } \right|^8 \right \}
\left \{ \left( \frac{\theta _3}{\eta } \right)^4 
+\left( \frac{\theta _4}{\eta } \right)^4 \right\}
 \nonumber \\ 
&
\hspace{3cm}
+
 \left| \frac{\theta _2}{\eta } \right| ^8
\left\{ 
\overline{\left( \frac{\theta _3}{\eta } \right)^4} +
\overline{\left( \frac{\theta _4}{\eta} \right)^4}  \right \}
\left \{ \left( \frac{\theta _3}{\eta } \right)^4 
+\left( \frac{\theta _4}{\eta } \right)^4 \right\},  \label{partitions without shift2}
\end{align}
Obviously, we cannot construct any left-moving supercharges since we find 
$$Z_{1,3}^{(\mathrm{NS} ,\mathrm{NS})}
(\tau , \bar \tau ) \neq - Z_{a}^{(\mathrm{R} ,\mathrm{NS})}
(\tau , \bar \tau ) , \hspace{1cm} 
( \any a \in \mathbb Z_4).
$$
%%%%
On the other hand, there would exist
some right supercharges in the $a=2$ sector which realizes 
the equalities 
\begin{equation}
Z_{a}^{(* ,\mathrm{NS})}
(\tau , \bar \tau ) = - Z_{a+2 \ \mathrm{mod}\ 4}^{(* ,\mathrm{R})}
(\tau , \bar \tau ),
\label{a=2 SUSY}
\end{equation}
as found  in 
\eqn{partitions without shift1}, \eqn{partitions without shift2}.
%%%%%%
In fact, one can explicitly confirm that the $a=2$ sector includes the right-moving massless  Ramond states,  
even though all of them are projected out by $(-1)^{F_R}|_{\psi}$ in the untwisted sector.
To be more precise, if starting with the type IIA (IIB) string theory, 
one can construct 8 supercharges that possess the {\em opposite chirality as those in the type IIB (IIA) theory\/}
from the $a=2$ sector, 
as discussed {\em e.g.} in \cite{Sen:1998ex,Gaberdiel:1999jd}.
%%%%

\begin{table}[h]
\begin{center}
	\caption{Massless spectrum in the $a=2$ sector for the orbifold model defined by 
$\sigma$
}
\label{mlspect0}
{\renewcommand\arraystretch{1.2}
\vspace{2mm}
\begin{tabular}{|c|c|}
\hline
spin structure & 4D fields\\
\hline
\hline
 (NS, R) & 2 gravitini, \\
&14 Weyl fermions\\
\hline 
 (R , R)  & 8 vectors, \\
  & 16 (pseudo)scalars \\
\hline
\end{tabular}}
\end{center}
\end{table}
%%%%%%%%%%%%%%%%%%%%%%%%%%%%%%%%%%%%%%%%%%%%%%%%%%%%%%%%%%%%%%%%%%%%%%%%%%
The massless spectrum in the untwisted sector is the same as that displayed
in Table \ref{mlspect1}, while that lying in the $a=2$ sector is summarized  in Table \ref{mlspect0}.
These states  are combined into the super-multiplets in an $N=2$ supersymmetric theory in 4-dimension.

~

%%%%%%%%%%%%%%%%%%%%%%%%%%%%%%%%%%%%%%%%%%%%%%%%%%%%%%%%%%%%%%%%%%%%%%%%
%%%%%%%%%%%%%%%%%%%%%%%%%%%%%%%%%%%%%%%%%%%%%%%%%%%%%%%%%%%%%%%%%%%%%%%%
%%%%%%%%%%%%%%%%%%%%%%%%%%%%%%%%%%%%%%%%%%%%%%%%%%%%%%%%%%%%%%%%%%%%%%%%
\section{Variety of Non-Supersymmetric Asymmetric Orbifolds }
\label{The other non-\susic asymmetric orbifolds }

In this section, we  present the main analyses in this paper.
As already mentioned in section 1, we especially focus on the modifications of the previous model introduced in section 2
 by {\bf (i)} replacing $(-1)^{F_L}|_{\psi}$ 
with $(\-_L)^{\otimes 2}$
in \eqn{g previous}, 
or/and  {\bf (ii)} requiring that 
$\cR \equiv (\-_R)^{\otimes 4}$ 
acts as the $\bz_4$-chiral reflection {\em also on the bosonic sector\/}
instead of including the shift  $\cT_{2\pi R}$.

We shall start our analyses with constructing the relevant building blocks in subsection \ref{building},
with emphasizing the modular covariance of them. 
%%%%
After that, we present the new six vacua composed of asymmetric orbifolds, and 
concretely discuss their physical aspects in subsection \ref{non-SUSY vacua}.
%%%%%%%%%
The readers not interested  in the technical part of this work may skip  many parts of subsection \ref{building}, and 
can refer only to the definitions of  building blocks.

~

%%%%%%%%%%%%%%%%%%%%%%%%%%%%%%%%%%%%%%%%%%%%%%%%%%%%%%%%%%%%%%%%%%%%%%
%------------------------------------------------------------------
\subsection{Building blocks}
\label{building}
\subsubsection{Bosonic $T^{N} [SO({2N})]$ Sector }
\label{block la}

Firstly, we discuss the simple example $T^2[SO(4)]$, identified as the $X^6$, $X^7$-directions. 
%that is, the 2-dim. torus at the $SO(4)$-point. 
The torus partition function of this system is 
\begin{align}
{Z}^{T^2[SO(4)]} (\tau ,\bar \tau) 
= \frac{1}{2}\left \{ \left| \frac{\theta _3}{\eta } \right|^4 
+ \left| \frac{\theta _4}{\eta } \right|^4
+\left| \frac{\theta _2}{\eta } \right|^4 \right \} \label{Z(SO4)}.
\end{align}
%%%%
A convenient description  is given by 
introducing the Majorana-Weyl fermions $\lambda ^i _{L}, \lambda ^i _{R}$
($i=1,2,3,4$). 
%%%%%

In the previous section, for simplicity, it has been assumed that 
the twist operator $\sigma $ including chiral reflection 
%$(-\mathbf 1_R)$ or $(-\mathbf 1_L)$ 
acts as an involution on the untwisted sector of the bosonic part.
However, once adopting the fermionic description of $T^2 [SO(4)]$, 
we are aware of another possibility  
in the manner similar to the  world-sheet fermions $\psi^{\mu}_L$, $\psi^{\mu}_R$. 
Namely, 
%let $\la_L^i$, $\la_R^i$ ($i=1,\ldots, 4$) be the free fermions describing $T^2 [SO(4)]$.
considering the left-mover for instance, the chiral reflection
$(-\mathbf 1_L)^{\otimes 2}|_{T^2}:(X^6_L, X^7_L) \to (-X^6_L, -X^7_L) $
is just interpretable as the sign flip of two of $\lambda ^1_ L, \ldots, \lambda ^4_L$, say, 
%%%%
%%%%
\begin{equation}
(-\mathbf 1_L)^{\otimes 2} : (\lambda_L ^1, \lambda_L ^2, \lambda_L ^3, \lambda_L ^4 )
\to (\lambda_L ^1, \lambda_L ^2, -\lambda_L ^3, -\lambda_L ^4 ).
\label{def cr la}
\end{equation}
%%%%
%%%%
As illustrated in \cite{SSW} and already mentioned in section 2 for the world-sheet fermions $\psi^{\mu}$, 
we still need to define the Ramond vacua of this free fermion system to  specify completely  
the action of $(\-_L)^{\otimes 2}$.
% on the Ramond vacua of $\la^i_L$. 
Here, there are essentially two different cases;
%%%
%%%%%%%
\begin{description}
\item[(a)\ $  \{(-\mathbf 1_L)^{\otimes 2}\} ^2= (-1)^{F_L}|_\lambda  $ :
]\mbox{}\\
%%%%%
One can introduce  the spin fields as 
\begin{equation}
\tS_{\epsilon _1, \epsilon _2, \, L }\equiv e^{i \sum^2_{i=1} \epsilon_i {\tH}^i_L}, \ \ \ \left(\epsilon _i=\pm \frac{1}{2}\right),
\end{equation}
with the bosonization;
\begin{equation}
\lambda ^1_L \pm \lambda ^2_L \equiv \sqrt 2e^{\pm i {\tH}^1_L},\ \ \ 
\lambda ^3_L \pm \lambda ^4_L \equiv \sqrt 2e^{\pm i {\tH}^2_L}.
\end{equation}
%%%%%
Then, \eqn{def cr la} yields 
\begin{equation}
(-\mathbf 1_L)^{\otimes 2} : (\tH_L^1, \tH_L^2)
\to (\tH^1_L, \tH^2_L+\pi ),
\end{equation}
and the Ramond vacua $\ket{\ep_1, \ep_2}_{L} \equiv \tS_{\epsilon _1, \epsilon _2, \, L}(0) \ket{0}_L$ are transformed as 
\begin{equation}
(-\mathbf 1_L)^{\otimes 2} \ket{\ep_1, \ep_2}_{L} = e^{i\pi \ep_2} \ket{\ep_1, \ep_2}_{L}.
\end{equation}
%%%%%
Thus, we find that  $ \{(-\mathbf 1_L)^{\otimes 2}\} ^2= -\mathbf 1$ holds for the R sector, 
while $ \{(-\mathbf 1_L)^{\otimes 2}\} ^2= \mathbf 1$ for the NS sector of $\la^i_L$.
Namely, we obtain $\{(-\mathbf 1_L)^{\otimes 2}\} ^2= (-1)^{F_L}|_\lambda  $.

%%%%%%%%%%%%%%%%%%%%%%%%%%%%%%%%%%%%%%%%%%%%%%%%%%%%%%%%%%%%%%%%%%%%

\item[(b)\ $  \{(-\mathbf 1_L)^{\otimes 2}\} ^2= \mathbf 1 $ :
]\mbox{}\\
One may  also bosonize $\lambda^1_L, \ldots, \la^4_R$ in a different way; 
\begin{equation}
\lambda ^1_L \pm \lambda ^3_L \equiv \sqrt 2e^{\pm i {\tH}^{' 1}_L},\ \ \ 
\lambda ^2_L \pm \lambda ^4_L \equiv \sqrt 2e^{\pm i {\tH}^{' 2}_L},
\end{equation}
and define the spin fields as follows;
\begin{equation}
\tS'_{\epsilon _1, \epsilon _2, \, L }\equiv e^{i \sum^2_{i=1} \epsilon_i {\tH}^{' i}_L}, \ \ \ \left(\epsilon _i=\pm \frac{1}{2} \right).
\end{equation}
This time,  \eqn{def cr la} yields 
\begin{equation}
(-\mathbf 1_L)^{\otimes 2} : ({\tH}_L^{' 1}, {\tH}_L^{' 2})
\to (-{\tH}^{'1}_L, -{\tH}^{' 2}_L),
\end{equation}
and the Ramond vacua 
$\ket{\ep_1, \ep_2}'_{L} \equiv S'_{\epsilon _1, \epsilon _2}(0) \ket{0}$
are transformed as 
\begin{equation}
(-\mathbf 1_L)^{\otimes 2} \ket{\ep_1, \ep_2}'_{L} = \ket{-\ep_1, -\ep_2}'_{L}.
\end{equation}
We thus simply obtain
$ \{(-\mathbf 1_L)^{\otimes 2}\} ^2= \mathbf 1 $.

\end{description}
%%%%%%%%%%%%%%%%%%%%%%%%%%%%%%%%%%%%%%%%%%%%%%%%%%%%%%%%%%%%%%%%%%%%%%%

The above arguments are straightforwardly generalized to the cases of 
$T^N[SO(2N)]$ ($N\in 2\bz_{>0}$) described by $2N$ Majorana-Weyl fermions $\la^i_L, ~ \la_R^i$ $(i=1, \ldots, 2N)$,
and we always have two possibilities; 
{\bf (i)} $  \{(-\mathbf 1_L)^{\otimes N}\} ^2= \mathbf 1 $, or {\bf (ii)} $  \{(-\mathbf 1_L)^{\otimes N}\} ^2= (-1)^{F_L}|_\lambda  $.

%%%%%%%%%%%%%%%%%%%%%%%%%%%%%%%%%%%%%%%%%%%%%%%%%%%%%%%%%%%%%%%%%%%%%%%%

~

Let us describe the relevant blocks which we will utilize later.
%%%%
In the following, the twist parameters $a,b \in \bz$ in 
the subscript always labels the spatial and temporal boundary conditions\footnote
   {Here, we shall allow the parameters of twisting $a,b$ to be arbitrary integers just for convenience, 
 although it is enough to restrict their range at most as $a,b \in \bz_4$.}. 
In other words, the parameter $a$ labels the twisted sectors, while 
the parameter $b$ corresponds to the insertion of $\sigma^b$ into the trace. 

%%%%

~

\noindent
%\underline
{\bf (i) $(-1)^{F_L}|_{\la}$-twisting in the $T^{N}[SO({2N})]$-sector :}

First we consider the building blocks $\mathbf{Z}^{T^{N}[SO({2N})]}_{(a,b)}(\tau , \bar \tau)$, $(a,b \in \bz)$, 
defined by the twisting $(-1)^{F_L}|_{\la}$ acting on $T^{N}[SO({2N})]$.
%%%%
The $(0,b)$-sector is just the insertion of $\left\{ (-1)^{F_L}|_{\la}\right\}^b$ into the trace, and 
easily  evaluated as 
\begin{equation}
\mathbf{Z}^{T^{N}[SO({2N})]}_{(0,b)}(\tau , \bar \tau)=
\left\{
\begin{array}{ll}
{Z}^{T^{N}[SO({2N})]} (\tau ,\bar \tau), & ~~ (b\in 2\bz),
\\
\frac{1}{2}\left \{ \left| \frac{\theta _3}{\eta } \right|^{2N}
+ \left| \frac{\theta _4}{\eta } \right|^{2N}
-\left| \frac{\theta _2}{\eta } \right|^{2N} \right \}, 
& ~~ (b \in 2\bz+1).
\end{array}
\right.
\end{equation}
%%%%
Then, requiring the modular covariance
\begin{align}
\mathbf{Z}^{T^{N}[SO({2N})]}_{(a,b)}(\tau , \bar \tau)|_S
=\mathbf{Z}^{T^{N}[SO({2N})]}_{(b,-a)}(\tau , \bar \tau ) ,
\nonumber \\
\mathbf{Z}^{T^{N}[SO({2N})]}_{(a,b)}(\tau , \bar \tau)|_T
=\mathbf{Z}^{T^{N}[SO({2N})]}_{(a,a+b)}(\tau , \bar \tau), 
\label{mod cov}
\end{align} 
we obtain 
\begin{align}
\mathbf{Z}^{T^{N}[SO({2N})]}_{(a,b)}(\tau ,\bar \tau) 
\equiv \left \{ 
\begin{array}{cl}
\frac{1}{2}\left \{ \left| \frac{\theta _3}{\eta } \right|^{2N} 
+ \left| \frac{\theta _4}{\eta } \right|^{2N}
+\left| \frac{\theta _2}{\eta } \right|^{2N} \right \} ,
& 
(a\in 2\mathbb Z, ~ b\in 2\mathbb Z) 
\\
\frac{1}{2}\left \{ \left| \frac{\theta _3}{\eta } \right|^{2N}
+ \left| \frac{\theta _4}{\eta } \right|^{2N}
-\left| \frac{\theta _2}{\eta } \right|^{2N} \right \} ,
& (a\in 2\mathbb Z, ~ b\in 2\mathbb Z+1) \\
\frac{1}{2}\left \{ \left| \frac{\theta _3}{\eta } \right|^{2N} 
- \left| \frac{\theta _4}{\eta } \right|^{2N}
+\left| \frac{\theta _2}{\eta } \right|^{2N} \right \} ,
& (a\in 2\mathbb Z +1,~ b\in 2\mathbb Z) \\
\frac{1}{2}\left \{ -\left| \frac{\theta _3}{\eta } \right|^{2N} 
+ \left| \frac{\theta _4}{\eta } \right|^{2N}
+\left| \frac{\theta _2}{\eta } \right|^{2N} \right \} , 
& (a\in 2\mathbb Z+1, ~ b\in 2\mathbb Z+1) . \\
\end{array}
\right . \label{defE(SON)}
\end{align}

~

%%%%%%%%%%%%%%%%%%%%%%%%%%%%%%%%%%%%%%%%%%%%%%%%%%%%%%%%%%%

\noindent
{\bf (ii) $(-\mathbf 1_R)^{\otimes N}$-twisting in the $T^{N}[SO({2N})]$-sector : }

Next, we consider the building blocks corresponding to the 
twist operator $(-\mathbf 1_R)^{\otimes 4}$
which acts on $T^4[SO(8)]$.
Yet, the twist operator is not specified.
%%%%%%
As noticed at the beginning of this section,
we have two possibilities  $\{(-\mathbf 1_R)^{\otimes 4}\} ^2=\mathbf 1 $, or $\{(-\mathbf 1_R)^{\otimes 4}\} ^2=(-1)^{F_R} |_{\la} $.

The building blocks for the first case 
are given as follows;
\begin{align}
&\lefteqn{{F}^{T^4[SO(8)]}_{(a,b)} (\tau ,\bar \tau) } 
\nonumber
\\
&
\hspace{1cm}
\equiv \left \{ 
\begin{array}{cl}
\frac{1}{2}\left \{ \left| \frac{\theta _3}{\eta } \right|^8 
+ \left| \frac{\theta _4}{\eta } \right|^8
+\left| \frac{\theta _2}{\eta } \right|^8 \right \} ,
& (a\in 2\mathbb Z, b\in 2\mathbb Z) \\
(-1)^{\frac{a}{2}} \overline{\left( \frac{\theta _3\theta _4}{\eta ^2} \right)^2}
\frac{1}{2}
\left \{ \left( \frac{\theta _3}{\eta } \right)^4 
+\left( \frac{\theta _4}{\eta } \right)^4 \right\},
& (a\in 2 \mathbb Z, b \in 2 \mathbb Z +1) \\
(-1)^{\frac{b}{2}} \overline{\left( \frac{\theta _2\theta _3}{\eta ^2} \right)^2}
\frac{1}{2}
\left \{ \left( \frac{\theta _3}{\eta } \right)^4 
+\left( \frac{\theta _2}{\eta } \right)^4 \right\} ,
& (a \in 2\mathbb Z +1, b\in 2\mathbb Z) \\
e^{-\frac{i\pi }{2}ab} \overline{\left( \frac{\theta _2\theta _4}{\eta ^2} \right)^2}
\frac{1}{2}
\left \{ \left( \frac{\theta _4}{\eta } \right)^4 
-\left( \frac{\theta _2}{\eta } \right)^4 \right\} ,
& (a \in 2\mathbb Z +1, b\in 2\mathbb Z +1) .
\end{array}
\right . \label{defF(SO8)}
\end{align}

%%%%%%%%%%%%%%%%%%%%%%%%%%%%%%%%%%%%%%%%%%%%%%%%%%%%%%%%%%%%%%

On the other hand, 
The building blocks corresponding to 
$\{(-\mathbf 1_R)^{\otimes 4}\} ^2= (-1)^{F_R}|_\lambda  $ 
are obtained by combining \eqn{defF(SO8)} with \eqn{defE(SON)};
%%%%%%%%%%%%%%%%%%%%%%%%%%%%%%%%%%%%%%%%%%%%%%%%%%%%%%%%%%%%%%%%%%%%%%%%%%%55
\begin{align}
%\lefteqn{{\mathbf F}^{T^4[SO(8)]}_{(a,b)} (\tau ,\bar \tau) }
{\mathbf F}^{T^4[SO(8)]}_{(a,b)} (\tau ,\bar \tau)
% \nonumber
%\\
%&
\equiv \left \{ 
\begin{array}{cl}
\mathbf Z^{T^4[SO(8)]}_{(\frac{a}{2},\frac{b}{2})}(\tau , \bar \tau),
& (a\in 2\mathbb Z, b\in 2\mathbb Z), \\
& \\
{F}^{T^4[SO(8)]}_{(a,b)} (\tau ,\bar \tau), 
& (a\in 2\mathbb Z +1\  \mathrm{or}\  b\in 2\mathbb Z +1) .\\
\end{array}
\right . \label{defFF(SO8)}
\end{align}
%%%%%%%%%%%%%%%%%%%%%%%%%%%%%%%%%%%%%%%%%%%%%%%%%%%%
Similarly, the building blocks for the 
$
 (-\mathbf 1_R)^{\otimes 2}%|_{T^2[SO(4)]}
$
-twisting 
on $T^2[SO(4)]$
are written as 
\begin{align}
\lefteqn{{F}^{T^2[SO(4)]}_{(a,b)} (\tau ,\bar \tau) } \nonumber
\\
&
\hspace{1cm}
\equiv \left \{ 
\begin{array}{cl}
\frac{1}{2}\left \{ \left| \frac{\theta _3}{\eta } \right|^4
+ \left| \frac{\theta _4}{\eta } \right|^4
+\left| \frac{\theta _2}{\eta } \right|^4 \right \} ,
& (a\in 2\mathbb Z, b\in 2\mathbb Z) \\
e^{\frac{i\pi }{4}ab} \overline{\left( \frac{\theta _3\theta _4}{\eta ^2} \right)}
\frac{1}{2}
\left \{ \left( \frac{\theta _3}{\eta } \right)^2 
+(-1)^{\frac{a}{2}}\left( \frac{\theta _4}{\eta } \right)^2 \right\},
& (a\in 2 \mathbb Z, b \in 2 \mathbb Z +1) \\
e^{-\frac{i\pi }{4}ab} \overline{\left( \frac{\theta _2\theta _3}{\eta ^2} \right)}
\frac{1}{2}
\left \{ \left( \frac{\theta _3}{\eta } \right)^2 
+(-1)^{\frac{b}{2}} \left( \frac{\theta _2}{\eta } \right)^2 \right\} ,
& (a \in 2\mathbb Z +1, b\in 2\mathbb Z) \\
e^{-\frac{i\pi }{4}ab} \overline{\left( \frac{\theta _2\theta _4}{\eta ^2} \right)}
\frac{1}{2}
\left \{ \left( \frac{\theta _4}{\eta } \right)^2 
-i(-1)^{\frac{a+b}{2}}\left( \frac{\theta _2}{\eta } \right)^2 \right\} ,
& (a \in 2\mathbb Z +1, b\in 2\mathbb Z +1) ,
\end{array}
\right . \label{defF(SO4)}
\end{align}
and,   
for the case of $\{(-\mathbf 1_R)^{\otimes 2}\} ^2= (-1)^{F_R}|_\lambda  $,  
%----------------------------------------------------------------------
\begin{align}
%\lefteqn{{\mathbf F}^{T^2[SO(4)]}_{(a,b)} (\tau ,\bar \tau) } \nonumber
%\\
{\mathbf F}^{T^2[SO(4)]}_{(a,b)} (\tau ,\bar \tau)
\equiv \left \{ 
\begin{array}{cl}
\mathbf{Z}^{T^2[SO(4)]}_{(\frac{a}{2},\frac{b}{2})}(\tau , \bar \tau),
& (a\in 2\mathbb Z, b\in 2\mathbb Z), \\
& \\
{F}^{T^2[SO(4)]}_{(a,b)} (\tau ,\bar \tau),
& (a\in 2\mathbb Z +1\  \mathrm{or}\  b\in 2\mathbb Z +1) .\\
\end{array}
\right . \label{defFF(SO4)}
\end{align}

%%%%%%%%%%%%%%%%%%%%%%%%%%%%%%%%%%%%%%%%%%%%%%%%

~

\noindent
{\bf (iii) twisting by  $(\-_L) \otimes (\-_R)^{\otimes 3}$ : }

For the later convenience, 
we also consider  the building blocks corresponding to 
twisting 
\begin{equation}
 \sigma \equiv 
(-\mathbf 1_L)|_{X^6}\otimes (-\mathbf 1_R)^{\otimes 3}|
_{X^{7,8,9}}, 
\end{equation}
acting on $T^4[SO(8)]$.
They are obtained in the same way as above. Namely, we first evaluate the 
trace with the twist operator inserted, and then all the building blocks are uniquely determined 
by requiring  the modular covariance such as \eqn{mod cov}.
The explicit computation  is straightforward, but  a little more cumbersome 
%to determine 
about the phase factors 
than those for the blocks $F^{T^N[SO(2N)]}_{(a,b)}$ given above. 
They are summarized as 
%%%%
\begin{align}
&{G}^{T^{4}[SO({8})]}_{(a,b)} (\tau ,\bar \tau) \nonumber \\
&\equiv \left \{ 
\begin{array}{cl}
\frac{1}{2}\left \{ \left| \frac{\theta _3}{\eta } \right|^{{8}} 
+ \left| \frac{\theta _4}{\eta } \right|^{{8}}
+\left| \frac{\theta _2}{\eta } \right|^{{8}} \right \} ,
& (a\in 2\mathbb Z, b\in 2\mathbb Z) ,\\
e^{\frac{i\pi }{4}ab} \overline{\left( \frac{\theta _3\theta _4}{\eta ^2} \right)}
\left| \frac{\theta _3\theta _4}{\eta ^2} \right| 
\frac{1}{2}
\left \{ \left( \frac{\theta _3}{\eta } \right)^2 
\left| \frac{\theta _3}{\eta} \right| ^{2}
+(-1)^{\frac{a}{2}}\left( \frac{\theta _4}{\eta } 
\right)^2
\left| \frac{\theta _4}{\eta } \right| ^{2}
\right\} ,
& (a\in 2 \mathbb Z, b \in 2 \mathbb Z +1), \\
e^{-\frac{i\pi }{4}ab} \overline{\left( \frac{\theta _2\theta _3}{\eta ^2} \right)}
\left| \frac{\theta _2\theta _3}{\eta ^2} \right| 
\frac{1}{2}
\left \{ 
\left( \frac{\theta _3}{\eta } \right)^2 
\left| \frac{\theta _3}{\eta } \right| ^{2}
+(-1)^{\frac{b}{2}}\left( \frac{\theta _2}{\eta } \right)^2 
\left| \frac{\theta _2}{\eta } \right| ^{2}
\right\},
& (a \in 2\mathbb Z +1, b\in 2\mathbb Z) ,\\
e^{-\frac{i\pi }{4}ab} \overline{\left( \frac{\theta _2\theta _4}{\eta ^2} \right)}
\left| \frac{\theta _2\theta _4}{\eta ^2} \right| 
\frac{1}{2}
\left \{ \left( \frac{\theta _4}{\eta } \right)^2
\left| \frac{\theta _4}{\eta } \right| ^{2} 
-i (-1)^{\frac{a+b}{2}}\left( \frac{\theta _2}{\eta } \right)^2 
\left| \frac{\theta _2}{\eta } \right| ^{2}
\right\},
& (a \in 2\mathbb Z +1, b\in 2\mathbb Z +1) ,
\end{array}
\right . \label{defG(SO8)}
\end{align}
%----------------------------------------------------------------------
or
\begin{align}
{\mathbf G}^{T^4[SO(8)]}_{(a,b)} (\tau ,\bar \tau) 
\equiv \left \{ 
\begin{array}{cl}
\mathbf Z^{T^4[SO(8)]}_{(\frac{a}{2},\frac{b}{2})} (\tau ,\bar \tau),
 & (a\in 2\mathbb Z, b\in 2\mathbb Z) ,  \\
 & \\
{\mathbf G}^{T^4[SO(8)]}_{(a,b)} (\tau ,\bar \tau) ,
& (a\in 2\mathbb Z+1\ \mathrm{or}\ b\in 2\mathbb Z +1) ,
\end{array}
\right .
\label{defGG(SO8)}
\end{align}
in the case that the twist operator is not involutive in the untwisted sector\footnote
  {Stated more precisely, we have the four possibilities;
  (i) $\sigma^2={\bf 1}$, (ii) $\sigma^2=(-1)^{F_L}|_{\la}$, (iii) $\sigma^2=(-1)^{F_R}|_{\la}$,  (iv)  $\sigma^2=(-1)^{F_L+F_R}|_{\la}$.
However, since the spin structure of  $\la^i$ is diagonal, the cases (ii) and (iii) lead us to the same building blocks \eqn{defGG(SO8)},
while the case (iv) yields \eqn{defG(SO8)} as well as the case (i).  
}.

~

%%%%%%%%%%%%%%%%%%%%%%%%%%%%%%%%%%%%%%%%%%%%%%%%%%%%%%%%%%%%%%%%%%%%%%%%%%
%%%%%%%%%%%%%%%%%%%%%%%%%%%%%%%%%%%%%%%%%%%%%%%%%%%%%%%%%%%%%%%%%%%%%%%%%%

%------------------------------------------------------------------------
\subsubsection{Fermionic Sector }
%\subsubsection*{Fermion Partition Function}

We next consider the fermionic sector.
We first recall that 
the fermionic part of the partition function of the type II string 
on 10-dim. flat background is just written as 
\begin{align}
{Z}^{\psi ,\tilde \psi}_{\mathrm{type II} } (\tau ,\bar \tau) 
=\frac{1}{4}|\mathcal J(\tau ) |^2,
\label{10flatF}
\end{align}
where
\begin{align}
\mathcal J(\tau )\equiv
   \left( \frac{\theta _3}{\eta } \right)^4
- \left( \frac{\theta _4}{\eta } \right)^4
- \left( \frac{\theta _2}{\eta } \right)^4 \left(\equiv 0 \right).       
\label{J}
\end{align}
Its modular property is easily seen as\footnote
  {The equations \eqn{JtraT}, \eqn{JtraS} or the modular covariance relations  \eqn{mod cov f}
would look slightly subtle since we know $\cJ(\tau) \equiv 0 $. 
See {\em e.g.} \cite{SSW} for more rigid statements.   
 }  
\begin{align}
\mathcal J(\tau )|_T & = - e^{\frac{\pi i}{3}} \mathcal J (\tau ) ,
\label{JtraT}
\\
\ \ \ \mathcal J(\tau )|_S &=   \mathcal J (\tau ),       
\label{JtraS}
\end{align}
and thus,  \eqref{10flatF} is modular invariant. 
%%%%%%%%%%%%%%%%%%%%%%%%%%%%%%%%%%%%%%%%%%%%%%%%%%%%%%
The desired free fermion chiral blocks are given 
by making the suitable modifications of $\cJ(\tau)$
caused by the orbifold twists so as to be compatible with 
the modular invariance.

We present the relevant chiral blocks from now on. 
We only focus on the left-mover, and the right-mover is completely parallel.
Although the cases (i) and (ii) are already given {\em e.g.} in \cite{SSW}, 
we dare to present them for the convenience to readers.

%%%%%%%%%%%%%%%%%%%%%%%%%%%%%%%%%%%%%%%%%%%%%%%%%%%%%%%
%%%%%%%%%%%%%%%%%%%%%%%%%%%%%%%%%%%%%%%%%%%%%%%%%%%%%%%

%~

\newpage

%\subsubsection*{Chiral blocks}
%\par 

\noindent
{\bf (i) twisting by $(-1)^{F_L}|_{\psi}$ : }

We first describe the twisting by $(-1)^{F_L}|_{\psi}$ and denote the corresponding chiral blocks 
as $h_{(a,b)}(\tau )$.
Again it is easiest to first compute $h_{(0,b)}(\tau)$, which just means 
the insertion of $\left\{ (-1)^{F_L}|_{\psi}\right\}^b$ into the trace;
\begin{align}
h_{(0,b)} (\tau )= \left \{ 
\begin{array}{cl}
\mathcal J(\tau ),
& ( b\in 2\mathbb Z), \\
& \\
   \left( \frac{\theta _3}{\eta } \right)^4
- \left( \frac{\theta _4}{\eta } \right)^4
+ \left( \frac{\theta _2}{\eta } \right)^4 
& ( b \in 2 \mathbb Z +1). \\
\end{array}
\right . \label{defJJ0b}
\end{align}
Requiring the modular covariance
\begin{align}
\left [ h _{(a,b)}(\tau )\overline{\mathcal J(\tau )} \right] |_S
=\left [ h_{(b,-a)} (\tau )\overline{\mathcal J(\tau )} \right] , 
% \label{JJtraS}
\nonumber
\\
\left [ h _{(a,b)}(\tau )\overline{\mathcal J(\tau ) }\right] |_T
=\left [ h _{(a,a+b)}(\tau )\overline{\mathcal J(\tau )} \right] ,     
%\label{JJtraT}
\label{mod cov f}
\end{align}
we obtain
\begin{align}
h_{(a,b)} (\tau ) = \left \{ 
\begin{array}{cl}
\mathcal J(\tau ),
& (a\in 2\mathbb Z, ~ b\in 2\mathbb Z), \\
   \left( \frac{\theta _3}{\eta } \right)^4
- \left( \frac{\theta _4}{\eta } \right)^4
+ \left( \frac{\theta _2}{\eta } \right)^4 ,
& (a\in 2 \mathbb Z, ~ b \in 2 \mathbb Z +1) ,\\
   \left( \frac{\theta _3}{\eta } \right)^4
+ \left( \frac{\theta _4}{\eta } \right)^4
- \left( \frac{\theta _2}{\eta } \right)^4,
& (a \in 2\mathbb Z +1, ~ b\in 2\mathbb Z), \\
-\left \{    \left( \frac{\theta _3}{\eta } \right)^4
+ \left( \frac{\theta _4}{\eta } \right)^4
+ \left( \frac{\theta _2}{\eta } \right)^4 \right\}, 
& (a \in 2\mathbb Z +1, ~ b\in 2\mathbb Z +1) .
\end{array}
\right . \label{defJJ}
\end{align}
We note that the left-chiral blocks have to 
give rise to
the phase $-e^{-\frac{\pi i}{3}}$ under the T-transformation
to satisfy the modular covariance relation \eqn{mod cov f}.
%%%%%%%%%%%%%%%%%%%%%%%%%%%%%%%%%%%%%%%%%%%%%%%%%%%%%%%%%%%%%%%%%%%
$h_{(a,b)}$ $(a \in 2\mathbb Z+1$, or $ b \in 2\mathbb Z +1 )$ are
non-vanishing, which implies the SUSY breaking in the left-moving sector.

~

%%%%%%%%%%%%%%%%%%%%%%%%%%%%%%%%%%%%%%%%%%%%%%%%%%%%%%%%%%%%%%%%%
\noindent
{\bf (ii) twisting by $(-\mathbf 1_L)^{\otimes 4}$ : }

Next, we look at the chiral blocks
defined by the chiral reflection 
%$(-\mathbf 1_L)^{\otimes 4}$ 
\begin{align}
&(-\mathbf 1_L)^{\otimes 4}: 
\ (\psi _L^6, \psi _L^7, \psi_L^8, \psi _L^9 ) \to (-\psi _L^6, -\psi _L^7, -\psi_L^8, -\psi_L^9 ). 
%\\
%& \{ (-\mathbf 1_L)^{\otimes 4}\}^2=\mathbf 1. 
\label{as1}
\end{align} 
As illustrated  in \cite{SSW}, we have again two possibilities; 
$\{ (-\mathbf 1_L)^{\otimes 4}\}^2=\mathbf 1$, or 
$\{ (-\mathbf 1_L)^{\otimes 4}\}^2= (-1)^{F_L}|_{\psi}$.
%%%
We denote the chiral blocks for the first case as $f_{(a,b)}(\tau)$.
One can similarly determine them by computing $f_{(0,b)}(\tau) $ first, and requiring the modular covariance 
such as \eqn{mod cov f}. 
They are summarized as 
\begin{align}
f_{(a,b)} (\tau) = \left \{ 
\begin{array}{cl}
\mathcal J(\tau ),
& (a\in 2\mathbb Z, ~ b\in 2\mathbb Z), \\
%& 
%\\
e^{\frac{i\pi }{2}ab}
\left \{ \left( \frac{\theta _3}{\eta } \right)^2\left( \frac{\theta _4}{\eta } \right)^2  
-\left( \frac{\theta _4}{\eta } \right)^2 \left( \frac{\theta _3}{\eta } \right)^2 +0 \right\}, 
& (a\in 2 \mathbb Z, ~ b \in 2 \mathbb Z +1) ,\\
e^{\frac{i\pi }{2}ab}
\left \{ \left( \frac{\theta _3}{\eta } \right)^2\left( \frac{\theta _2}{\eta } \right)^2  
+0-\left( \frac{\theta _2}{\eta } \right)^2 \left( \frac{\theta _3}{\eta } \right)^2 \right\} ,
& (a \in 2\mathbb Z +1, ~ b\in 2\mathbb Z), \\
-e^{\frac{i\pi }{2}ab}
\left \{ 0+\left( \frac{\theta _2}{\eta } \right)^2\left( \frac{\theta _4}{\eta } \right)^2  
-\left( \frac{\theta _4}{\eta } \right)^2 \left( \frac{\theta _2}{\eta } \right)^2 \right\} ,
& (a \in 2\mathbb Z +1, ~ b\in 2\mathbb Z +1).
\end{array}
\right.  \label{deff(a,b)}
\end{align}
Note that all of them trivially vanish as is consistent with 
the preservation of half space-time SUSY in the left-mover. 
%%%%%
Each term from the left to the right corresponds to the spin structures;  NS, $\tNS$, 
and R sector, respectively,
where the `$\tNS$' denotes the NS-sector with 
$(-1)^f$ inserted 
($f$ is the world-sheet fermion number).
%%%%
%%%%%%%%%%%%%%%%%%%%%%%%%%%%%%%%%%%%%%%%%%%%5

On the other hand, 
in the second case $\{ (-\mathbf 1_L)^{\otimes 4}\}^2=(-1)^{F_L}|_\psi $,
the relevant chiral blocks are just modified as follows;
\begin{align}
\mathbf f _{(a,b)} (\tau )\equiv \left \{ 
\begin{array}{cl}
h_{(\frac{a}{2},\frac{b}{2})}(\tau ),
& (a\in 2\mathbb Z, ~ b\in 2\mathbb Z), \\
& \\
f _{(a,b)} (\tau ),
& (a \in 2\mathbb Z +1\ \mathrm{or}\ b\in 2\mathbb Z +1) .
\end{array}
\right . \label{defff(a,b)a}
\end{align}
Recall that $h_{(*,*)}(\tau)$ is given in \eqn{defJJ}, corresponding to $(-1)^{F_L}|_{\psi}$-twisting.

~

%%%%%%%%%%%%%%%%%%%%%%%%%%%%%%%%%%%%%%%%%%%%%%%%%%%%%%%%%%%%%%

\noindent
{\bf (iii) twisting by $(-\mathbf 1_L)^{\otimes 2}$ : }

We also need the chiral blocks defined by 
$(-\mathbf 1_L)^{\otimes 2}$-twisting. 
They are determined in the parallel way as above, although 
the different phase factors have to be included to ensure the modular covariance.

For the case 
$\{ (-\mathbf 1_L)^{\otimes 2}\}^2=\mathbf 1$, we obtain 
\begin{align}
g_{(a,b)} (\tau )\equiv \left \{ 
\begin{array}{cl}
\mathcal J(\tau ),
& (a\in 2\mathbb Z, b\in 2\mathbb Z), \\
& \\
e^{-\frac{i\pi }{4}ab}
\left \{ \left( \frac{\theta _3}{\eta } \right)^3\left( \frac{\theta _4}{\eta } \right)  
-(-1)^{\frac{a}{2}}\left( \frac{\theta _4}{\eta } \right)^3 \left( \frac{\theta _3}{\eta } \right) +0 \right \} ,
& (a\in 2 \mathbb Z, b \in 2 \mathbb Z +1) ,\\
e^{\frac{i\pi }{4}ab}
\left \{ \left( \frac{\theta _3}{\eta } \right)^3\left( \frac{\theta _2}{\eta } \right)  
+0
-(-1)^{\frac{b}{2}}\left( \frac{\theta _2}{\eta } \right)^3 \left( \frac{\theta _3}{\eta } \right) \right \} ,
& (a \in 2\mathbb Z +1, b\in 2\mathbb Z), \\
-e^{\frac{i\pi }{4}ab}
\left \{ 0+\left( \frac{\theta _4}{\eta } \right)^3\left( \frac{\theta _2}{\eta } \right)  
+i(-1)^{\frac{a+b}{2}}\left( \frac{\theta _2}{\eta } \right)^3 \left( \frac{\theta _4}{\eta } \right) \right \} ,
& (a \in 2\mathbb Z +1, b\in 2\mathbb Z +1) ,
\end{array}
\right.  
\label{defg(a,b)}
\end{align}
and 
\begin{align}
\mathbf g _{(a,b)} (\tau )\equiv \left \{ 
\begin{array}{cl}
h_{(\frac{a}{2},\frac{b}{2})}(\tau ),
& (a\in 2\mathbb Z, b\in 2\mathbb Z), \\
& \\
g _{(a,b)} (\tau ),
& (a \in 2\mathbb Z +1\ \mathrm{or}\ b\in 2\mathbb Z +1) ,
\end{array}
\right.  \label{defgg(a,b)a}
\end{align}
for 
$\{ (-\mathbf 1_L)^{\otimes 2}\}^2= (-1)^{F_L}|_{\psi}$.

~

%%%%%%%%%%%%%%%%%%%%%%%%%%%%%%%%%%%%%%%%%%%%%%%%%%%
%%%%%%%%%%%%%%%%%%%%%%%%%%%%%%%%%%%%%%%%%%%%%%%%%%%
%%%%%%%%%%%%%%%%%%%%%%%%%%%%%%%%%%%%%%%%%%%%%%%%%%%

\subsection{Non-\susic Asymmetric orbifolds} 
\label{non-SUSY vacua}

We are now ready to study 
the six new models of non-SUSY vacua 
exhibited in Tables \ref{7twists} and \ref{7backgrounds}, 
including the modifications introduced 
at the beginning of this section.

~

%\subsubsection*{Asymmetric Orbifold I}
\subsubsection*{Model I :}

Firstly, we consider the asymmetric orbifold defined by the orbifold twist 
%\eqref{asym actionI}
%\begin{align}
\begin{eqnarray}
g 
&=& \mathcal T_{2\pi R}|_{\mathrm{base}}\otimes
% \sigma_\mathrm{I}|_{T^4}
\sigma_{\msc{I}} 
\equiv \mathcal T_{2\pi R}|_{\mathrm{base}}
\otimes
 (-\mathbf 1)^{\otimes 2} |_{X^{6,7}}
\otimes (-\mathbf 1_R)^{\otimes 2}|_{X^{8,9}}
\nonumber
\\
&\equiv&  \mathcal T_{2\pi R}|_{\mathrm{base}}
\otimes
 (-\mathbf 1_L)^{\otimes 2} |_{X^{6,7}}
\otimes (-\mathbf 1_R)^{\otimes 4}|_{X^{6,7,8,9}},
\label{asym actionI}
\end{eqnarray}
%\end{align}
acting on 
\begin{equation}
\left[ M^{4}\times S^1 \right] \times \mathbb R
_{\mathrm{base}} 
\times T^2_{6,7}
\times T^2_{8,9}
%_{\mathrm{fiber}}
[SO(4)] . 
\label{bgI}
\end{equation}
%where the second $T^2$ from the right is an arbitrary two dimensional torus. 
In the above expressions we explicitly indicated 
the directions along which the orbifold twist \eqn{asym actionI} acts in terms of  the subscripts.
%%%%
Namely, 
the $X^{8,9}$-directions are compactified on $T^2[SO(4)]$, while the $X^{6,7}$-directions 
correspond to a 2-dim. torus with unspecified moduli. 
Note that the non-chiral reflection $(\-)^{\otimes 2} \, : \, (X^6,X^7) \, \longmapsto \, (-X^6, -X^7)$
is well-defined for any point of moduli space of $T^2$.
$\cT_{2\pi R}$ denotes the shift by $2\pi R$  along the $\br_{\msc{base}}$.
%%%%

As addressed in section 2, we further need to specify  the Ramond vacua of world-sheet fermions and the action of $\sigma_{\msc{I}}$ on them. 
%Since $\sigma_I$ non-trivially acts both on the right and left-moving fermions,    
Adopting the Ramond vacua defined by the bosonization given in \eqn{bosonization} both for the right and left movers,  
we can naturally define 
%%%%%%%%%%%%%%%%%%%%%%%%%%%%
\begin{align}
& \sigma_{\msc{I}} \ket{s_1,s_2,s_3,s_4}_R = 
e^{ i\pi s_4} \ket{s_1,-s_2,-s_3,s_4}_R,
\nonumber\\
& \sigma_{\msc{I}} \ket{s_1,s_2,s_3,s_4}_L = 
\ket{s_1,-s_2,-s_3,s_4}_L.
\label{def sigmaI right f}
\end{align}
which  implies 
\begin{equation}
\sigma_{\msc{I}} ^2 = (-1)^{F_R} |_{\psi}.
\end{equation}
%%%%%%%%%%%%%%%%%%%%%%%%%%%%
%%%%%%%%%%%%%%%%%%%%%%%%%%%%

We can write down 
the torus partition function in terms of the building blocks introduced in subsection \ref{building} as
\begin{equation}
Z(\tau , \bar \tau )=
\frac{1}{4}{\mathcal Z}_{M^4\times S^1}
(\tau ,\bar \tau)
\sum_{w,m \in \mathbb Z}
Z_{R, (w,m)}(\tau ,\bar \tau )
{Z}^{T^2/\mathbb Z_2}_{(w,m)}(\tau,\bar \tau )
{F}^{T^2[SO(4)]}_{(w,m)}(\tau,\bar \tau )
{g}_{(w,m)}(\tau) 
\overline{\mathbf f_{(w,m)}(\tau)}.
\label{partitionI}
\end{equation}
%Here and in the subsequent models, we shall simply  denote 
As in section 2, we simply denote
the contributions with no relations to the orbifolding  
as `${\mathcal Z_*}$'.
In the current case, 
${\mathcal Z}_{M^4\times S^1}$ is identified as 
that for the bosonic transverse part of $M^4\times S^1$-sector ($X^{0, \ldots , 4}$-directions).
$Z_{R, (w,m)}(\tau ,\bar \tau )$ is
% the familiar partition function of the winding sector of the compact boson with radius $R$
given in  \eqn{Rblock},
%%%%%%%%%
%%%%%%%%%
while
${Z}^{T^2/\mathbb Z_2}_{(w,m)}(\tau,\bar \tau )$ 
%($w,m \in \bz_2$) 
expresses  
the building blocks of the symmetric $\bz_2$-orbifold along the $X^{6,7}$-directions.
(We have an obvious $\bz_2$-periodicity with respect to the winding $w$, $m$.)
The bosonic building blocks 
${F}^{T^2[SO(4)]}_{(w,m)}(\tau,\bar \tau )$ are given in 
\eqn{defF(SO4)},
while the  chiral blocks for world-sheet fermions, denoted  
as ${g}_{(w,m)}(\tau) $,
$\mathbf f_{(w,m)}(\tau)$, 
are presented  in \eqn{defg(a,b)}, 
\eqn{defff(a,b)a},  respectively.
Looking at their expressions, it is easy to 
confirm that the partition function \eqref{partitionI} indeed  vanishes
in the manner similar to the arguments in section 2.

%%%%%%%%%%%%%%%%%%%%%%%%%%%%%%%%%%%%%%%%%%%%%%%%%%%%%%%%%%%%%%%%%%%%%

As noticed in section 1, the non-SUSY chiral reflection $(\-_L)^{\otimes 2}$
plays the similar role of $(-1)^{F_L}|_{\psi}$ in the `previous model'  introduced in section 2, 
and thus we anticipate to achieve a non-SUSY vacuum  with the bose-fermi cancellation. 
We will later show that this is indeed the case.

%%%%%%%%%%%%%%%%%%%%%%%%%%%%%%%%%%%%%%%%%%%%%%%%%%%%%%%%%%%%%%%%

Before doing so, 
let us study the massless spectrum lying in the untwisted sector, which we  summarize  
in Table \ref{mlI}. We express the left-moving Ramond vacua 
in terms of the spin fields for $SO(8)$;  $\ket{\mathbf s}_L\equiv e^{i\sum_{a=1}^4 \, s_a H_L^a}  \ket{0}_L, ~
\left(s_a \equiv \pm \frac{1}{2}
\right)
$.
%%%%%%%%%%%%%%%%%%%%%%%%%%%%%%%%%%%
\begin{table}[ht]
\begin{center}
	\caption{Massless spectrum in untwisted sector for asymmetric orbifold I}
\label{mlI}
{\renewcommand\arraystretch{1.3}
\vspace{2mm}
\begin{tabular}{|c|ccc|c|}
\hline
spin structure & left & & right & 4D fields (d.o.f)\\
\hline
\hline
 &$\psi ^\mu _{L, -1/2}\ket 0 $& $\otimes $ & 
$ \psi ^\mu  _{R, -1/2}\ket 0 $& graviton (2), 6 vectors (12),\\ 
 (NS, NS) & $(\mu =2,...,7)$ & & $(\mu =2,...,5) $ & 10 (pseudo) scalars (10)\\ \cline{2-5}
 &$\psi ^\mu _{L, -1/2}\ket 0 $& $\otimes $ & 
$ \psi ^\mu  _{R, -1/2}\ket 0 $& 8 scalars (8) \\
 & $(\mu =8,9)$ & & $(\mu =6,...,9) $ &  \\
\hline 
  &
$\left[1 + ({\bf -1}_L)^{\otimes 2}\right] \ket{\mathbf {s}}_L$
& $\otimes $ & 
$\psi ^\mu  _{R, -1/2}\ket 0  $ & 2 Rarita-Schwinger (4),  \\
 (R , NS) &  & & $(\mu =2,...,5)$  & 6 Weyl fermions (12)\\ \cline{2-5}
  &
%$\ket{\mathbf s_1} - \ket{\mathbf {s_2}}$
$\left[1 - ({\bf -1}_L)^{\otimes 2}\right] \ket{\mathbf {s}}_L$
& $\otimes $ & 
$\psi ^\mu  _{R, -1/2}\ket 0  $ & 8 Weyl fermions (16) \\
 &  & & $(\mu =6,...,9)$  & \\
\hline
\end{tabular}
}
\end{center}
\end{table}
%%%%%%%%%%%%%%%%%%%%

What is a remarkable difference from the previous model is the existence of 
massless Rarita-Schwinger fields. 
They of course originate from the gravitini in the original background \eqn{bgI},
which are not removed by the relevant orbifold projection. 
In the same sense, some supercharges in the original background remain preserved under 
the orbifold group.

Nonetheless, the space-time SUSY is completely broken within the untwisted sector,
at least. It is obvious not to have the right-moving supercharges due to the absence of 
right-moving Ramond vacua. 
Furthermore, even though having the left-moving Ramond vacua, 
we  cannot still compose any left-moving supercharges acting as isomorphisms 
on the orbifolded  Hilbert space.
%in the orbifold theory. 
In fact, the presence of left supercharges should
imply the existence of one to one correspondence between the (NS,NS) 
and (R,NS) massless states, {\it while fixing the right-movers}.
It is, however, impossible as shown  from Table \ref{mlI}.
For instance,  pick up  the states $ \psi ^\mu  _{R, -1/2}\ket 0, \  (\mu =6,...,9)$ from the
right-mover. Then, one finds that 
the 
%number of 
degrees of freedom of massless bosons 
%((NS,NS)-states)
amount to  8,  whereas the fermionic one 
%((R,NS)-states) 
is 16.

%%%%%%%%%%%%%%%%%%%%%%%%%%%%%%%%%%%%%%%%%%%%%%%%%%%%%%%%%%%%%%%%%%%%

One can examine the more detailed spectrum of physical states 
by making the Poisson resummation of the partition function \eqref{partitionI}.
To this aim
it is convenient to decompose it with respect to
the spatial winding $w$ and the spin structures as in \eqref{departition0};
\begin{align}
& Z(\tau , \bar \tau )=\frac{1}{4}{\mathcal Z}
_{M^4\times S^1}
(\tau ,\bar \tau)
\nonumber \\
& \hspace{1cm} \times
\sum _{w \in \mathbb Z} \left \{
 Z_w^{(\mathrm{NS}, \mathrm{NS})}(\tau , \bar \tau )
+Z_w^{(\mathrm{NS}, \mathrm{R})}(\tau , \bar \tau )
+Z_w^{(\mathrm{R}, \mathrm{NS})}(\tau , \bar \tau )
+Z_w^{(\mathrm{R}, \mathrm{R})}
(\tau , \bar \tau )\right \}      .
 \label{departition}
\end{align}
After dualizing 
%Poisson resumming 
the temporal winding $m$ into the KK momentum $n$, 
we obtain the following results;
%%%%%%%%%%%%%%%%%%%%%%%%%%%%%%%%%%%%%%%%%%%%%
\begin{itemize}

\item
For $w\in 4\mathbb Z$:
\begin{align}
& 
\lefteqn{Z_w^{(\mathrm{NS} ,\mathrm{NS})}
(\tau , \bar \tau ) = -Z_w^{(\mathrm{R} ,\mathrm{NS})}
(\tau , \bar \tau ) } \label{ZNSNSw4ZI} \\
&
\hspace{1cm}
=
\frac{1}{2} \sum_{n\in \mathbb Z} \frac{1}{|\eta |^2}
q^{\frac{1}{4}\left( R w +\frac{n}{2R} \right )^2}
\bar q^{\frac{1}{4}\left( R w -\frac{n}{2R} \right )^2}
Z^{T^2}Z^{T^2[SO(4)]} 
\left | \left( \frac{\theta _3}{\eta} \right )^4 
-\left( \frac{\theta _4}{\eta} \right )^4  \right |^2 ,
\nonumber \\
%\end{align}
%\begin{align}
& \lefteqn{Z_w^{(\mathrm{R} ,\mathrm{R})}
(\tau , \bar \tau ) = -Z_w^{(\mathrm{NS} ,\mathrm{R})}
(\tau , \bar \tau)} 
\nonumber \\
&
\hspace{1cm}
=
\frac{1}{2}
\sum_{n\in \mathbb Z} \frac{1}{|\eta |^2}
q^{\frac{1}{4}\left( R w +\frac{n+\frac{1}{2}}{2R} \right )^2}
\bar q^{\frac{1}{4}\left( R w -\frac{n+\frac{1}{2}}{2R} \right )^2}
Z^{T^2}Z^{T^2[SO(4)]} 
\left| \frac{\theta _2}{\eta} \right |^8 . 
 \label{Z4ZOb}
\end{align}

%%%%%%%%%%%%%%%%%%%%%%%%%%%%%%%%%%%%%%%%%%%%%%%%%%%%%%%%%%%%%%%%%%%%%%%

\item $w\in 4\mathbb  Z +2$:
\begin{align}
&Z_w^{(\mathrm{NS} ,\mathrm{NS})}
(\tau , \bar \tau ) = -Z_w^{(\mathrm{R} ,\mathrm{NS})}
(\tau , \bar \tau ) =
\frac{1}{2} \sum_{n\in \mathbb Z}\frac{1}{|\eta |^2}
q^{\frac{1}{4}\left( R w +\frac{n+\frac{1}{2}}{2R} \right )^2}
\bar q^{\frac{1}{4}\left( R w -\frac{n+\frac{1}{2}}{2R} \right )^2} 
\nonumber \\
& 
\hspace{1cm}
\times 
Z^{T^2}Z^{T^2[SO(4)]} 
\left \{ \left( \frac{\theta _3}{\eta} \right )^4 -\left( \frac{\theta _4}{\eta} \right )^4 \right \}
\left \{ \overline{ \left( \frac{\theta _3}{\eta} \right ) }^4 
+\overline{\left( \frac{\theta _4}{\eta} \right )}^4 \right \} ,
\label{ZNSNSw4Z2I}
\\
&
\lefteqn{Z_w^{(\mathrm{R} ,\mathrm{R})}
(\tau , \bar \tau ) = -Z_w^{(\mathrm{NS} ,\mathrm{R})}
(\tau , \bar \tau)}           
 \nonumber  \\
&
\hspace{1cm}
=\frac{1}{2} \sum_{n\in \mathbb Z}\frac{1}{|\eta |^2}
q^{\frac{1}{4}\left( R w +\frac{n}{2R} \right )^2}
\bar q^{\frac{1}{4}\left( R w -\frac{n}{2R} \right )^2}
Z^{T^2}Z^{T^2[SO(4)]} 
\left| \frac{\theta _2}{\eta} \right |^8  .  
\label{ZRRw4Z2I}
\end{align}

%%%%%%%%%%%%%%%%%%%%%%%%%%%%%%%%%%%%%%%%%%%%%%%

\item
For $w\in 4\mathbb Z +1$:
%%%%%%%%%%%%%%%%%%%%%%%%%%%
\begin{align}
& \hspace{-5mm}
Z_w^{(\mathrm{NS} ,\mathrm{NS})}
(\tau , \bar \tau ) = -Z_w^{(\mathrm{NS} ,\mathrm{R})}
(\tau , \bar \tau ) 
\nonumber \\
& 
%\hspace{5mm}
=\frac{1}{4} \sum _{a\in \mathbb Z_2 } 
\sum_{n\in \mathbb Z} \left[
%\frac{1}{2}
\frac{1}{|\eta |^2} q^{\frac{1}{4}\left( R w +\frac{n+\frac{1}{2}}{2R} \right )^2}
\bar q^{\frac{1}{4}\left( R w -\frac{n+\frac{1}{2}}{2R} \right )^2}
%\nonumber \\ &\times 
(-1)^{an} 
\left| \frac{\theta _2\theta _3(\frac{a}{2})}{\eta ^2} \right| ^4
%\frac{1}{2}
\left( \frac{\theta _3(\frac{a}{2})}{\eta } \right)^4
\overline{\left( \frac{\theta _3(\frac{a}{2})\theta _2}{\eta ^2} \right)}^2
%\label{ANSNSmI}\\
\right.
\nonumber \\
&
\hspace{2.5cm}
\left.
+
%\sum _{a\in \mathbb Z_2 } 
%\frac{1}{|\eta |^2}\sum_{n\in \mathbb Z}
%\frac{1}{2}
\frac{1}{|\eta |^2} 
q^{\frac{1}{4}\left( R w +\frac{n}{2R} \right )^2}
\bar q^{\ \frac{1}{4}\left( R w -\frac{n}{2R} \right )^2} (-1)^{an}
%\frac{1}{2} 
\left| \frac{\theta _2 \theta _3(\frac{a}{2})}{\eta ^2} \right|^8
\right] ,
\label{BNSNSmI}
\\
%\end{align}
%\begin{align}
%%%%%%%%%%%%%%%%%%%%%%%%%%
& \hspace{-5mm}
Z_w^{(\mathrm{R} ,\mathrm{R})}
(\tau , \bar \tau ) = -Z_w^{(\mathrm{R} ,\mathrm{NS})}
(\tau , \bar \tau ) 
\nonumber \\
& 
%\hspace{5mm}
=\frac{1}{4} \sum _{a\in \mathbb Z_2 } 
\sum_{n\in \mathbb Z} \left[
%\frac{1}{2}
\frac{1}{|\eta |^2} q^{\frac{1}{4}\left( R w +\frac{n+\frac{1}{2}}{2R} \right )^2}
\bar q^{\frac{1}{4}\left( R w -\frac{n+\frac{1}{2}}{2R} \right )^2}
%\nonumber \\ &\times 
(-1)^{a(n+1)} 
\left| \frac{\theta _2\theta _3(\frac{a}{2})}{\eta ^2} \right| ^4
%\frac{1}{2}
\left( \frac{\theta _2}{\eta } \right)^4
\overline{\left( \frac{\theta _2 \theta _3(\frac{a}{2})}{\eta ^2} \right)}^2
%\label{ANSNSmI}\\
\right.
\nonumber \\
&
\hspace{2.5cm}
\left.
+
%\sum _{a\in \mathbb Z_2 } 
%\frac{1}{|\eta |^2}\sum_{n\in \mathbb Z}
%\frac{1}{2}
\frac{1}{|\eta |^2} 
q^{\frac{1}{4}\left( R w +\frac{n}{2R} \right )^2}
\bar q^{\ \frac{1}{4}\left( R w -\frac{n}{2R} \right )^2} (-1)^{an}
%\frac{1}{2} 
\left| \frac{\theta _2 \theta _3(\frac{a}{2})}{\eta ^2} \right|^8
\right]  .
\label{BRRmI}
\end{align}

%%%%%%%%%%%%%%%%%%%%%%%%%%%%%

\item For $w\in 4\mathbb Z +3$:

In this case, the result is obtained by replacing $(-1)^{an}$ 
in the first term of \eqn{BNSNSmI}
%for (\ref{ANSNSmI}) 
with $(-1)^{a(n+1)}$, 
and by replacing  $(-1)^{a(n+1)}$ in the first term of \eqn{BRRmI}
%for (\ref{ARRmI}) 
with $(-1)^{an}$.

%%%%%%%%%%%%%%%%%%%%%%%%%%%%%%%

\end{itemize}

%%%%%%%%%%%%%%%%%%%%%%%%%%%%%%%%%%%%%%%%%%%%%%%%%%%%%%%%%%%%%%%%%%%

All of these partition functions are 
$q$-expanded so as to be compatible with unitarity, and we have no tachyonic states 
as confirmed by looking at the conformal weights read from  them.
%%%
Extra massless excitations appear when the $X^5$-direction
has some specific radii, as summarized in Table \ref{massless pointsI}.
Moreover, it is easy to confirm that the above partition functions  
satisfy the same relation as given in Table \ref{all partition relation}
with respect to the winding number $w$. This fact makes it clear that 
the model I is indeed 
a non-SUSY vacuum with the bose-fermi cancellation at each mass level.

%%%%%%%%%%%%%%%%%%%%%%%%%%%%%%%%%%%%%%%%%%%%%%%%%%%%%%%
%-----------------------------------------------------------
\begin{table}[ht]
\begin{center}
\caption{The massless points for asymmetric orbifold I}
\label{massless pointsI}
{\renewcommand\arraystretch{1.2}
\vspace{2mm}
\begin{tabular}{|c|c|c|c|}
\hline
spin structure & massless point & sector & relevant equation
\\
\hline
\hline
 (NS, NS) / (NS, R)
  &$ R=\frac{1}{2}$  & $w=\pm 1$ & 
\eqn{BNSNSmI}
%\eqref{ANSNSmI}
 \\   
\hline
 (NS, NS) / (R, NS)
  &$ R=\frac{1}{2\sqrt 2}$  & $w=\pm2$ & \eqref{ZNSNSw4Z2I}
\\
\hline

\end{tabular}
}
\end{center}
\end{table}
%%%%%%%%%%%%%%%%%%%%%%%%%%%%%%%%%%%%%%%%%%%%%%%%%%%%%%%%%%%%%%%%%%%%%
In Table \ref{massless pointsI} the `relevant equation' indicates which partition function
 includes the terms corresponding to the massless states  
in question.

~

%%%%%%%%%%%%%%%%%%%%%%%%%%%%%%%%%%%%%%%%%%%%%%%%%%%%%%%%%%%%%%%%%%%%%%%%%%%%%%%%%%%
%%%%%%%%%%%%%%%%%%%%%%%%%%%%%%%%%%%%%%%%%%%%%%%%%%%%%%%%%%%%%%%%%%%%%%%%%%%%%%%%%%%
%%%%%%%%%%%%%%%%%%%%%%%%%%%%%%%%%%%%%%%%%%%%%%%%%%%%%%%%%%%%%%%%%%%%%%%%%%%%%%%%%%%

%\subsubsection*{Asymmetric Orbifold II}

\subsubsection*{Model II :}
%%%

The model II is defined by the orbifold twist 
\begin{align}
{g}
&  =\mathcal T_{2\pi R}|_{\mathrm{base}}\otimes
\sigma_{\msc{II}}
 %\sigma_\mathrm{II}|_{T^5}
 \equiv  \mathcal T_{2\pi R}|_{\mathrm{base}}
\otimes (-\mathbf 1)|_{X^5}
\otimes
 (-\mathbf 1_L) |_{X^{6}}
\otimes (-\mathbf 1_R)^{\otimes 3}|_{ X^{7,8,9}}, 
\nonumber
 \\
& \equiv 
\mathcal T_{2\pi R}|_{\mathrm{base}}\otimes
(-\mathbf 1_L)^{\otimes 2} |_{X^{5, 6}}
\otimes (-\mathbf 1_R)^{\otimes 4}|_{ X^{5, 7,8,9}},
\label{asym actionII}
\end{align}
acting on the background 
\begin{equation}
[M^{4} ] \times \mathbb R
_{\mathrm{base}} 
\times S_5^1
\times T^4_{6,7,8,9}
%_{\mathrm{fiber}}
[SO(8)] . \label{bgII}
\end{equation}
%%%%%%%%%%%%%%%%%%%%%%%%%%%%
%%%%%%%%%%%%%%%%%%%%%%%%%%%%
For the Ramond vacua, we set
\begin{align}
& \sigma_{\msc{II}} \ket{s_1,s_2,s_3,s_4}_R = 
e^{ i\pi s_4} \ket{s_1,-s_2,-s_3,s_4}_R,
\nonumber\\
& \sigma_{\msc{II}} \ket{s_1,s_2,s_3,s_4}_L = 
e^{i\pi s_3} \ket{s_1,-s_2,-s_3,s_4}_L.
\label{def sigmaII right f}
\end{align}
which again implies 
\begin{equation}
\sigma_{\msc{II}} ^2 = (-1)^{F_R} |_{\psi}.
\end{equation}
%%%%%%%%%%%%%%%%%%%%%%%%%%%%

The corresponding  partition function is given as 
\begin{equation}
Z(\tau , \bar \tau )=
\frac{1}{4}{\mathcal Z}_{M^{4}}
(\tau ,\bar \tau)
\sum_{w,m \in \mathbb Z}
Z_{R(w,m)}(\tau ,\bar \tau )
{Z}^{S^1/\mathbb Z_2}_{(w,m)}(\tau,\bar \tau )
{G}^{T^4[SO(8)]}_{(w,m)}(\tau,\bar \tau )
{g}_{(w,m)}(\tau) 
\overline{\mathbf f _{(w,m)}(\tau)}, \label{partitionII}
\end{equation}
where ${Z}^{S^1/\mathbb Z_2}_{(w,m)}(\tau,\bar \tau )$ denotes 
the building blocks corresponding to the ordinary reflection
$-\mathbf 1 \, :\,  (X^5_L, X^5_R) \, \to \,  (-X^5_L, -X^5_R) $ acting on $S_5^1$ with an arbitrary radius.
The bosonic blocks 
${G}^{T^4[SO(8)]}_{(w,m)}(\tau,\bar \tau )$ are defined in
\eqn{defG(SO8)}.

%%%%%%%%%%%%%%%%
This model is quite similar to the model I, although the partition function is slightly
different.
The massless spectrum and the massless points for the winding states 
are the same as that of the model I.
This result is independent of the radius of the $S_5^1$.
%%%%%%%%%%%%%%%%

~

%%%%%%%%%%%%%%%%%%%%%%%%%%%%%%%%%%%%%%%%%%%%%%%%%%%%%%%%%%%%%
%%%%%%%%%%%%%%%%%%%%%%%%%%%%%%%%%%%%%%%%%%%%%%%%%%%%%%%%%%%%%
%%%%%%%%%%%%%%%%%%%%%%%%%%%%%%%%%%%%%%%%%%%%%%%%%%%%%%%%%%%%%

%\subsubsection*{Asymmetric Orbifold III}

\subsubsection*{Model III :}
%%%

From now on, we shall  discuss the constructions of non-SUSY vacua without the  shift operator
$\cT_{2\pi R}$ included.
The simplest case, which we call  model III, is defined on  the background
\begin{equation}
[M^{4}\times T^2]
\times T^4_{6,7,8,9}
%_{\mathrm{fiber}}
[SO(8)] , \label{bgIII}
\end{equation}
and the orbifold twist is obtained simply as  
\begin{equation}
{g} = \sigma'
%\sigma |_{T^4}
\equiv
(-1)^{F_L}|_\psi 
\otimes (-\mathbf 1_R)^{\otimes 4}|_{X^6,\ldots , X^9}.
\label{asym actionIII}
\end{equation}
%%%%%%%%%%%%%%%%%%%%%
Although it looks almost the same as the supersymmetric vacua illustrated 
%at \eqn{sample partition'} 
in section 2, we shall here adopt the $\bz_4$-action 
as the definition of $(-\mathbf 1_R)^{\otimes 4}|_{X^6,\ldots , X^9}$ {\em also for the bosonic sector\/}
by utilizing the fermionization as given in subsection \ref{block la}.
Namely, 
introducing the free fermions $\la^i_{L \, (R)}$, ($i=1,\ldots, 8$) describing 
$T^4[SO(8)]$, we identify $(-\mathbf 1_R)^{\otimes 4}|_{X^6,\ldots , X^9}$ with the sign flip of $\la_R^5, \ldots, \la_R^8$.
We then  determine  its action on the Ramond vacua of $\la^i_R$ as  
\begin{align}
(-\mathbf 1_R)^{\otimes 4}|_{X^6,\ldots , X^9} ~ : ~ \ket{\ep_1,\ep_2,\ep_3,\ep_4}_{\la, R} ~ \longmapsto ~ e^{i\pi \ep_4} \ket{\ep_1,-\ep_2,-\ep_3,\ep_4}_{\la, R},
\end{align}
with the definitions 
%%%%%%%%%%%%%%%%%%%%%%%%%%%
\begin{align}
& \ket{\ep_1,\ep_2,\ep_3,\ep_4}_{\la, L \,(R)} 
%= \tS_{\ep_1, \ep_2, \ep_3, \ep_4, \, L \, (R)} \ket{0}_{\la, L\, (R)} 
\equiv 
e^{i\sum_{a=1}^4 \ep_a \tH^a_{L \, (R)}} \ket{0}_{\la, L \, (R)},  \hspace{1cm} \left( \ep_a = \pm \frac{1}{2}\right),
\nonumber \\
& \la_{L \, (R)}^1 \pm i\la_{L \, (R)}^2 = \sqrt{2} e^{\pm i \tH_{L \, (R)}^1}, ~~~ \la_{L \, (R)}^3 \pm i\la_{L \, (R)}^5 = \sqrt{2} e^{\pm i \tH_{L \, (R)}^2}, 
\nonumber
\\
&
\la_{L \, (R)}^4 \pm i\la_{L \, (R)}^6 = \sqrt{2} e^{\pm i \tH_{L \, (R)}^3}, ~~~ \la_{L \, (R)}^7 \pm i\la_{L \, (R)}^8 = \sqrt{2} e^{\pm i \tH_{L \, (R)}^4}. 
\label{bosonization la SO(8)}
\end{align}
%%%%%%%%%%%%%%%%%%%%%%%%%%%%
We also assume that $(-\mathbf 1_R)^{\otimes 4}|_{X^6,\ldots , X^9}$ acts on the Ramond vacua of the world-sheet fermions $\psi^{\mu}_R$
in the same way as \eqn{def sigma right f}.
In total, we obtain  
\begin{equation}
\left(\sigma'\right)^2 = (-1)^{F_R}|_{\lambda} \otimes (-1)^{F_R}|_{\psi} ,
\label{sigmaIII2}
\end{equation}
rather than 
$
\sigma^2 = (-1)^{F_R}|_{\psi}.
$
%%%%%%%%%%%%%%%%%%%%%%%%%%%%%%%%%%%%%%%%%%%%%%%%%%%%%%%%%%%%%%%%%%%

As we emphasized before, the shift operator $\cT_{2\pi R} $ plays an important role of SUSY breaking,
that is, it prevents the twisted sectors from providing new supercharges.
However, we here show that other types of non-SUSY vacua are realized 
%without the shift operator
as long as  \eqn{sigmaIII2} is satisfied.

The partition function is just written as
\begin{equation}
Z(\tau , \bar \tau )=\frac{1}{16}{\mathcal Z}_{M^4\times T^2}
(\tau ,\bar \tau )\sum_{a,b \in \mathbb Z_4}
\mathbf{F}^{T^4[SO(8)]}_{(a,b)}(\tau,\bar \tau )
h_{(a,b)}(\tau) \overline{\mathbf f_{(a,b)}(\tau)},
\label{partitionIII}
\end{equation}
where $\mathbf{F}^{T^4[SO(8)]}_{(a,b)}(\tau,\bar \tau )$, $h_{(a,b)}(\tau) $ and 
${\mathbf f_{(a,b)}(\tau)}$ are presented respectively in 
\eqn{defFF(SO4)},
\eqn{defJJ}, and 
\eqn{defff(a,b)a}. 
This partition function \eqn{partitionIII} also vanishes as is readily checked.

Let us  decompose
% the partition function 
\eqn{partitionIII}
with respect to  the twisted sectors as well as the spin structures
as 
\begin{align}
& Z(\tau , \bar \tau )=\frac{1}{16}{\mathcal Z_{M^4 \times T^2}}(\tau ,\bar \tau)
\nonumber
\\
& \hspace{1cm}
\times
\sum _{a \in \mathbb Z_4} \left \{
 Z_a^{(\mathrm{NS}, \mathrm{NS})}(\tau , \bar \tau )
+Z_a^{(\mathrm{NS}, \mathrm{R})}(\tau , \bar \tau )
+Z_a^{(\mathrm{R}, \mathrm{NS})}(\tau , \bar \tau )
+Z_a^{(\mathrm{R}, \mathrm{R})}
(\tau , \bar \tau )\right \} .           
 \label{departitionIII}
\end{align}
%%%%%%%%%%%%%%%%%%%%%%%%
Then,  we obtain
\begin{align}
&Z_0^{(\mathrm{NS} ,\mathrm{NS})}
(\tau , \bar \tau ) = -Z_0^{(\mathrm{R} ,\mathrm{NS})}
(\tau , \bar \tau )
=\left \{ \left|\frac{\theta _3}{\eta}\right |^8
+ \left|\frac{\theta _4}{\eta}\right|^8  \right \}
\left |\frac{\theta _2}{\eta} \right |^8 ,
\nonumber
\\
&Z_2^{(\mathrm{NS} ,\mathrm{NS})}
(\tau , \bar \tau ) = -Z_2^{(\mathrm{R} ,\mathrm{NS})}
(\tau , \bar \tau )
 \nonumber \\
&
\hspace{2cm}
=\left \{ \left|\frac{\theta _3}{\eta}\right |^8
- \left|\frac{\theta _4}{\eta}\right|^8  \right \}
\left (\frac{\theta _2}{\eta} \right )^4
\left \{ \overline{ \left(\frac{\theta _3}{\eta}\right )^4}
+ \overline{ \left( \frac{\theta _4}{\eta}\right) ^4 } \right \} ,
\nonumber
\\
%\end{align}
%\begin{align}
&Z_{0,2}^{(\mathrm{R} ,\mathrm{R})}
(\tau , \bar \tau ) = -Z_{0,2}^{(\mathrm{NS} ,\mathrm{R})}
(\tau , \bar \tau)
=
\left |\frac{\theta _2}{\eta} \right |^{16},
 \label{even III}  
\end{align}
for the even sectors, and 
\begin{align}
&Z_{1,3}^{(\mathrm{NS} ,\mathrm{NS})}
(\tau , \bar \tau ) =
 -Z_{1,3}^{(\mathrm{NS} ,\mathrm{R})}
(\tau , \bar \tau )
%\nonumber \\
%& ~~~
=
\overline{\left( \frac{\theta _2}{\eta } \right)^4} 
\left\{ 
\left| \frac{\theta _3}{\eta } \right|^8
-\left| \frac{\theta _4}{\eta } \right|^8 \right \}
\left \{ \left( \frac{\theta _3}{\eta } \right)^4 
+\left( \frac{\theta _4}{\eta } \right)^4 \right\},
 \nonumber \\ 
& \hspace{2cm}
+
 \left| \frac{\theta _2}{\eta } \right| ^8
\left\{ 
\overline{\left( \frac{\theta _3}{\eta } \right)^4} +
\overline{\left( \frac{\theta _4}{\eta} \right)^4}  \right \}
\left \{ \left( \frac{\theta _3}{\eta } \right)^4 
+\left( \frac{\theta _4}{\eta } \right)^4 \right\} , 
\nonumber
\\
%\end{align}
%\begin{align}
& Z_{1,3}^{(\mathrm{R} ,\mathrm{R})}
(\tau , \bar \tau ) = -Z_{1,3}^{(\mathrm{R} ,\mathrm{NS})}
(\tau , \bar \tau )
=\left \{ \left|\frac{\theta _3}{\eta}\right |^8
+ \left|\frac{\theta _4}{\eta}\right|^8  + \left |\frac{\theta _2}{\eta} \right |^8 \right \}
\left |\frac{\theta _2}{\eta} \right |^8,
%\nonumber
 \label{odd III}  
\end{align}
for the odd sectors.

%%%%%%%%%%%%%%%%%%%%%%%%%%%%%%%%%%%%%%%
These relations should be compared with those for the supersymmetric case 
\eqn{partitions without shift1} and \eqn{partitions without shift2}. 
Here we never  obtain the equalities  
such as \eqn{a=2 SUSY}, rather find the cancellations  as depicted in 
Table \ref{all partition relation}.
Namely, we see that the left-moving NS-R cancellations for the even sectors, whereas 
the right-moving ones for the odd sectors.
This fact clearly shows that the space-time SUSY is completely broken.
%%%
Recall that, 
in the supersymmetric case with $g=\sigma$, 
the right-moving SUSY is unbroken, and the supercharges arise from the  $a=2$ twisted sector.
In the current case, however, the same does not happen because the partition functions $Z^{(*, \msc{R})}_2(\tau,\bar{\tau}) $ 
do not contain any massless states. 
This is the crucial difference caused by the relation \eqn{sigmaIII2}.
In this way, we have successfully achieved a  desired non-SUSY vacuum without the shift. 

%%%

The massless spectrum in the untwisted sector is the same as the model introduced 
in the previous section. 
In the twisted sectors, on the other hand, 
there are additional massless states, while no tachyonic states appear.

~

%%%%%%%%%%%%%%%%%%%%%%%%%%%%%%%%%%%%%%%%%%%%%%%%%%%%%%%%%%%%%%%%%%%%%%%%%%%%
%--------------------------------------------------------------------------
%\subsubsection*{Asymmetric Orbifold IV}
\subsubsection*{Model IV :}
%%%

%Let us again try to construct a non-SUSY vacuum without the shift operator. 
We next consider the background
\begin{equation}
\left[ M^{4}\times T^2 \right] \times T_{6,7}^2
\times T_{8,9}^2
%_{\mathrm{fiber}}
[SO(4)] , \label{bgIV}
\end{equation}
and adopt the modification of \eqref{asym actionI}; 
\begin{align}
& g =\sigma'_{\msc{I}} \equiv (-\mathbf 1)^{\otimes 2} |_{X^{6,7}}
\otimes (-\mathbf 1_R)^{\otimes 2}|_{X^{8,9}}
%\nonumber
%\\
%& 
\equiv  (-\mathbf 1_L)^{\otimes 2} |_{X^{6,7}}
\otimes (-\mathbf 1_R)^{\otimes 2}|_{X^{6,7,8,9}},
%\nonumber
%\\
%& \left(\sigma'_{\msc{I}}\right)^2 = (-1)^{F_R}|_{ \lambda} \otimes (-1)^{F_R}|_{ \psi},
\end{align}
as the relevant orbifold twisting. 
%%%%%%%%%%%%%%%%%%%%%%%%%%%%%%%%%%%%%%%%%%%%%%%%%%
%%%%%%%%%%%%%%%%%%%%%%%%%%%%%%%%%%%%%%%%%%%%%%%%%%
$\sigma'_{\msc{I}}$ again acts by \eqn{def sigmaI right f} for the Ramond vacua of world-sheet fermions 
$\psi^{\mu}_R$, $\psi^{\mu}_L$.
On the other hand, introducing the free fermions $\la_{L \, (R)}^i$, ($i=1,\ldots, 4$)
describing $T^2_{8,9}[SO(4)]$, we define its action on the Ramond vacua of $\la^i_R$ as 
that given in {\bf (a)} of subsection \ref{block la}, that is,  
%%%%%%%%%%%%%%%%%%%%%%
\begin{align}
\sigma'_I \ket{\ep_1,\ep_2}_{\la, R} =  e^{i\pi \ep_2} \ket{\ep_1,\ep_2}_{\la, R},
\end{align}
with 
%%%%%%%%%%%%%%%%%%%%%%%%%%%
\begin{align}
& \ket{\ep_1,\ep_2}_{\la, R} 
%= \tS_{\ep_1, \ep_2, \, R} \ket{0}_{\la, R} 
\equiv 
e^{i\sum_{a=1}^2 \ep_a \tH^a_R} \ket{0}_{\la, R},  \hspace{1cm} \left( \ep_a = \pm \frac{1}{2}\right),
\nonumber \\
& \la_{L \, (R)}^1 \pm i\la_{L \, (R)}^2 = \sqrt{2} e^{\pm i \tH_{L \, (R)}^1}, ~~~ \la_{L \, (R)}^3 \pm i\la_{L \, (R)}^4 = \sqrt{2} e^{\pm i \tH_{L \, (R)}^2}.
\label{bosonization la SO(4)}
\end{align}
%%%%%%%%%%%%%%%%%%%%%%%%%%%%
We thus  obtain the crucial relation 
$
\left(\sigma'_{\msc{I}}\right)^2 = (-1)^{F_R}|_{ \lambda} \otimes (-1)^{F_R}|_{ \psi}.
$

%%%%%%%%%%%%%%%%%%%%%%%%%%%%%%%%%%%%%%%%%%%%%%%%%%
%%%%%%%%%%%%%%%%%%%%%%%%%%%%%%%%%%%%%%%%%%%%%%%%%%

The partition function is then written as 
\begin{equation}
Z(\tau , \bar \tau )=\frac{1}{16}{\mathcal Z}_{ M^4 \times T^2}
(\tau ,\bar \tau )\sum_{a,b \in \mathbb Z_4}
Z^{T^2/\mathbb Z_2}_{(a,b)}(\tau,\bar{\tau}) 
\mathbf{F}^{T^2[SO(4)]}_{(a,b)}(\tau,\bar \tau )
g_{(a,b)}(\tau) \overline{\mathbf f_{(a,b)}(\tau)},
\label{partitionIV}
\end{equation}
where 
$ \mathbf{F}^{T^2[SO(4)]}_{(a,b)}(\tau,\bar \tau )$, 
$ g_{(a,b)}(\tau)$, and $\mathbf f_{(a,b)}(\tau)$ are given respectively by 
\eqn{defFF(SO4)}, \eqn{defg(a,b)} and \eqn{defff(a,b)a}.
$Z^{T^2/\mathbb Z_2}_{(a,b)}(\tau,\bar{\tau})$ denotes 
the  building blocks corresponding to an ordinary  $\mathbb Z_2$-orbifold 
for the reflection acting $T_{6,7}^2$ in \eqref{bgIV}.
This partition function  also vanishes, and the supersymmetry is completely broken
at least in the untwisted sector, as confirmed in the same way as the case of model I.

The spectrum of physical states in each twisted sector is read off from the decomposition of 
partition function in the manner similar to
\eqn{departitionIII}. After a short computation, 
one finds the relations of the partition functions between 
all of the sectors as follows;
\begin{align}
%----
& Z_{0}^{(\mathrm{NS} ,\mathrm{NS})}
(\tau , \bar \tau ) = -Z_{0}^{(\mathrm{R} ,\mathrm{NS})}
(\tau , \bar \tau )
% \nonumber  \\
=Z^{T^2}
\left \{ \left|\frac{\theta _3}{\eta}\right |^4
+ \left|\frac{\theta _4}{\eta}\right|^4  \right \}
\left |\frac{\theta _2}{\eta} \right |^8, 
\nonumber \\
& Z_{0,2}^{(\mathrm{R} ,\mathrm{R})} (\tau , \bar \tau ) 
= -Z_{0,2}^{(\mathrm{NS} ,\mathrm{R})}
(\tau , \bar \tau )
=Z^{T^2}\left|\frac{\theta _2}{\eta}\right |^{12}, 
\nonumber
\\
%-----
& Z_{2}^{(\mathrm{NS} ,\mathrm{NS})}
(\tau , \bar \tau ) = -Z_{2}^{(\mathrm{R} ,\mathrm{NS})}
(\tau , \bar \tau ) 
%\nonumber  \\
=Z^{T^2}
\left \{ \left|\frac{\theta _3}{\eta}\right |^4
- \left|\frac{\theta _4}{\eta}\right|^4  \right \}
\left (\frac{\theta _2}{\eta} \right )^4
\left \{ \overline{ \left(\frac{\theta _3}{\eta}\right )^4}
+ \overline{ \left( \frac{\theta _4}{\eta}\right) ^4 } \right \} , 
%\nonumber
\label{even IV}
\\
%%%%%%%
%& Z_{2}^{(\mathrm{R} ,\mathrm{R})} (\tau , \bar \tau ) 
%= -Z_{2}^{(\mathrm{NS} ,\mathrm{R})}
%(\tau , \bar \tau )
%=Z^{T^2}\left|\frac{\theta_2}{\eta}\right|^{12},
%\nonumber
%\\
%%%%%
& Z_{1,3}^{(\mathrm{NS} ,\mathrm{NS})}
(\tau , \bar \tau ) = -Z_{1,3}^{(\mathrm{NS} ,\mathrm{R})}
(\tau , \bar \tau )
=-Z_{1,3}^{(\mathrm{R} ,\mathrm{NS})}
(\tau , \bar \tau ) = Z_{1,3}^{(\mathrm{R} ,\mathrm{R})}
(\tau , \bar \tau )
\nonumber
\\
&
\hspace{2.5cm} 
=\left|\frac{\theta _2\theta _3}{\eta ^2}\right |^6 Z^{T^2/\mathbb Z_2}_{(1,0)}
+ \left|\frac{\theta _2 \theta _4}{\eta ^2}\right|^6 Z^{T^2/\mathbb Z_2}_{(1,1)}
\equiv
\left|\frac{\theta _2}{\eta}\right |^ {8} \left\{ \left|\frac{\theta _3}{\eta }\right |^ 8
+ \left|\frac{\theta _4}{\eta }\right |^ 8 \right\}.
%\nonumber
 \label{odd IV}  
\end{align}
%%%%%%%%%%%%
These relations manifestly show that we do not have any right-moving supercharges. 
It is however curious that we have the `accidental' equalities  
%$Z^{\mathrm{(NS,NS)}}_a=-Z^{\mathrm{(R, NS)}}_a$
%and $Z^{\mathrm{(R,R)}}_a=-Z^{\mathrm{(NS,R)}}_a$  for $^\forall a \in \mathbb Z_4$
$$
Z^{\mathrm{(NS,NS)}}_a=-Z^{\mathrm{(R, NS)}}_a, ~~
Z^{\mathrm{(R,R)}}_a=-Z^{\mathrm{(NS,R)}}_a, ~~ (\mbox{for} ~ \any a \in \bz_4)
$$
in spite of the absence of left-moving supercharges 
that originate from the unorbifolded theory.
%in the untwisted sector.
It would be actually possible to make up some operators that realize these equalities 
%for, say, the odd sectors $a=1,3$,  
{\em by combining the spin fields of $\psi^{\mu}$ and $\la^i$}.
%since everything is written in terms of free fields. 
However, it turns out that such `fake' supercharges do not  
%describing $T^2[SO(4)]$, 
respect the super-Poincare symmetry in $M^4$ 
of the original background  \eqn{bgIV}.
%%%
Furthermore, any left-moving supercharges cannot be constructed also from the twisted sectors, because
we have
$$
Z^{\mathrm{(NS,NS)}}_a\neq -Z^{\mathrm{(R, NS)}}_{a'}, ~~
Z^{\mathrm{(R,R)}}_a\neq -Z^{\mathrm{(NS,R)}}_{a'}, ~~   (\mbox{for} ~ \any a'  \in \mathbb Z_4 ~ \mbox{s.t.} ~ a' \neq a).
$$

In this way, we conclude that the model IV is still a non-SUSY vacuum 
with the bose-fermi cancellation. 
Again we find additional massless states in the twisted sectors, while no tachyons appear.

~

%----------------------------------------------------------------------------

%%%%%%%%%%%%%%%%%%%%%%%%%%%%%%%%%%%%%%%%%%%%%%%%%%%%%%%%%%%%%%%%%%%%%%%%%%%%%%
%%%%%%%%%%%%%%%%%%%%%%%%%%%%%%%%%%%%%%%%%%%%%%%%%%%%%%%%%%%%%%%%%%%%%%%%%%%%%%

%\subsubsection*{Asymmetric Orbifold V}

\subsubsection*{Model V :}
%%%
The model V is defined similarly to the model IV on the background;
\begin{equation}
\left[ M^{4}\times S^1 \right]
\times S_5^1 
\times T_{6,7,8,9}^4[SO(8)],
\label{bgV}
\end{equation}
with the twist operator $g= \sigma'_{\msc{II}}$ which is the modification of  $\sigma_{\msc{II}}$ given in \eqn{asym actionII} as  
$\left(\sigma'_{\msc{II}}\right)^2= (-1)^{F_R}|_{\lambda} \otimes (-1)^{F_R}|_{\psi} $.
%%%%%%%%%%%%%%%%%%%%%%%%%%%%%%%%%%%%%%%%%%
%%%%%%%%%%%%%%%%%%%%%%%%%%%%%%%%%%%%%%%%%%
Namely, 
$\sigma_{\msc{II}}'$ acts on the world-sheet fermions $\psi^{\mu}_R$, $\psi^{\mu}_L$ in the same way as \eqn{asym actionII}, \eqn{def sigmaII right f},
and acts as the sign flip of 
$\la^5_L$, $\la^i_R$ ($i=6,7,8$), 
%$\la^6_R$, $\la^7_R$, $\la^8_R$
where $\la_{L \, (R)}^i$ ($i=1, \ldots, 8$) are the free fermions describing $T_{6,7,8,9}^4[SO(8)]$.
Moreover, its action on the Ramond vacua of $\la^i_R$, $\la^i_L$ 
is given as
\begin{align}
& \sigma_{\msc{II}}' \ket{\ep_1,\ep_2,\ep_3,\ep_4}_{\la, R} =  e^{i\pi \ep_4} \ket{\ep_1, \ep_2,-\ep_3,\ep_4}_{\la, R},
\nonumber 
\\
& \sigma_{\msc{II}}' \ket{\ep_1,\ep_2,\ep_3,\ep_4}_{\la, L} =   \ket{\ep_1, - \ep_2, \ep_3,\ep_4}_{\la, L},
\label{def sigmaII' la}
\end{align}
under the definitions \eqn{bosonization la SO(8)}.

%%%%%%%%%%%%%%%%%%%%%%%%%%%%%%%%%%%%%%%%%
%%%%%%%%%%%%%%%%%%%%%%%%%%%%%%%%%%%%%%%%%

The partition function is just given as 
\begin{equation}
Z(\tau , \bar \tau )=
\frac{1}{16}{\mathcal Z}_{M^4\times S^1}%^\mathrm{tr}_{M^4\times S^1}
(\tau ,\bar \tau)
\sum_{a,b \in \mathbb Z_4}
{Z}^{S^1/\mathbb Z_2}_{(a,b)}(\tau,\bar \tau )
\mathbf{G}^{T^4[SO(8)]}_{(a,b)}(\tau,\bar \tau )
{g}_{(a,b)}(\tau) 
\overline{\mathbf f _{(a,b)}(\tau)},
 \label{partitionV}
\end{equation}
where ${g}_{(a,b)}(\tau) $, $\mathbf f _{(a,b)}(\tau)$ are as above, while 
$\mathbf{G}^{T^4[SO(8)]}_{(a,b)}(\tau,\bar \tau )$ are presented in \eqn{defGG(SO8)}.

Since the fermionic building blocks are common with the model IV, 
we can likewise make the decomposition 
%of partition function 
such as \eqn{departitionIII};
\begin{align}
%----
&Z_{0}^{(\mathrm{NS} ,\mathrm{NS})}
(\tau , \bar \tau ) = -Z_{0}^{(\mathrm{R} ,\mathrm{NS})}
(\tau , \bar \tau ) 
%\nonumber  \\
%&
=Z^{S^1}
\left \{ \left|\frac{\theta _3}{\eta}\right |^8
+ \left|\frac{\theta _4}{\eta}\right|^8  \right \}
\left |\frac{\theta _2}{\eta} \right |^8,
\nonumber
\\
&Z_{0,2}^{(\mathrm{R} ,\mathrm{R})} (\tau , \bar \tau ) 
= -Z_{0,2}^{(\mathrm{NS} ,\mathrm{R})}
(\tau , \bar \tau )
=Z^{S^1}\left|\frac{\theta _2}{\eta}\right |^{16},
\nonumber
\\
%-----
&Z_{2}^{(\mathrm{NS} ,\mathrm{NS})}
(\tau , \bar \tau ) = -Z_{2}^{(\mathrm{R} ,\mathrm{NS})}
(\tau , \bar \tau ) ,
\nonumber  \\
&
\hspace{1.5cm}
=Z^{S^1}
\left \{ \left|\frac{\theta _3}{\eta}\right |^8
- \left|\frac{\theta _4}{\eta}\right|^8  \right \}
\left (\frac{\theta _2}{\eta} \right )^4
\left \{ \overline{ \left(\frac{\theta _3}{\eta}\right )^4}
+ \overline{ \left( \frac{\theta _4}{\eta}\right) ^4 } \right \} ,
%\\
\label{even V}
\end{align}
for the even sectors, and 
%%%
\begin{align}
&Z_{1,3}^{(\mathrm{NS} ,\mathrm{NS})}
(\tau , \bar \tau ) = -Z_{1,3}^{(\mathrm{NS} ,\mathrm{R})}
(\tau , \bar \tau ) 
=
\left|\frac{\theta _2}{\eta}\right |^ {10}\left\{ \left|\frac{\theta _3}{\eta }\right |^ 8
+ \left|\frac{\theta _4}{\eta }\right |^ 8 \right\},
\nonumber
\\
&Z_{1,3}^{(\mathrm{R} ,\mathrm{R})} (\tau , \bar \tau ) 
= -Z_{1,3}^{(\mathrm{R} ,\mathrm{NS})}
(\tau , \bar \tau )
=\left|\frac{\theta _2}{\eta}\right |^{8}\left\{ \left|\frac{\theta _3}{\eta }\right |^{10}
+\left|\frac{\theta _4}{\eta }\right |^{10}\right\},
 \label{odd V}  
\end{align}
for the odd sectors. 

The partition functions for even sectors \eqn{even V} coincide with  those for the model III \eqn{even III}
up to the common factor $Z^{S^1}$, implying that the right SUSY is completely broken. 
They are also quite similar to the model IV \eqn{even IV}, as anticipated.

Even though the odd sectors are also similar to \eqn{odd IV},  we here have a crucial difference. 
Namely, we find 
$
Z^{\mathrm{(NS,NS)}}_{1,3}\neq -Z^{\mathrm{(R, NS)}}_{1,3}, 
$ 
%in contrast with \eqn{odd IV}, 
just leading to the fact that the left SUSY is obviously broken.

The massless spectra in the twisted sectors are different from the previous two models.
For example, massless (NS,NS) and (NS, R) states do not appear in the odd  sectors as opposed to the case of model IV.
No tachyons appear as in the models so far.

~

%%%%%%%%%%%%%%%%%%%%%%%%%%%%%%%%%%%%%%%%%%%%%%%%%%%%%%%%%%%%%%%%%%%%

%\subsubsection*{Asymmetric Orbifold VI}

\subsubsection*{Model VI : }
%%%
Finally, we briefly mention on the model defined by the orbifold twist
$
(-\mathbf 1_L)^{\otimes 2} |_{X^{4,5}}
\otimes (-\mathbf 1_R)^{\otimes 4}|_{X^{6,7,8,9}},
$
%%%%
which is again organized to be a $\bz_4$-action both on the fermionic and bosonic sectors, 
%%%%
acting  on the background  
$
\left[ M^{4} \right] 
\times T^6_{4,5,\ldots, 9}[SO(12)] $. 
This is similar to the model V. 
The partition functions are almost the same as the previous two models;
\begin{align}
%-----
&Z_{0}^{(\mathrm{NS} ,\mathrm{NS})}
(\tau , \bar \tau ) = -Z_{0}^{(\mathrm{R} ,\mathrm{NS})}
(\tau , \bar \tau )
% \nonumber  \\
=
\left \{ \left|\frac{\theta _3}{\eta}\right |^{12}
+ \left|\frac{\theta _4}{\eta}\right|^{12} \right \}
\left |\frac{\theta _2}{\eta} \right |^8,
\nonumber
\\
&Z_{0}^{(\mathrm{R} ,\mathrm{R})} (\tau , \bar \tau ) 
= -Z_{0}^{(\mathrm{NS} ,\mathrm{R})}
(\tau , \bar \tau )
=\left|\frac{\theta _2}{\eta}\right |^{20},
\nonumber
 \\
%-----
&Z_{2}^{(\mathrm{NS} ,\mathrm{NS})}
(\tau , \bar \tau ) = -Z_{2}^{(\mathrm{R} ,\mathrm{NS})}
(\tau , \bar \tau ) 
%\nonumber  \\
%& 
=
\left \{ \left|\frac{\theta _3}{\eta}\right |^{12}
- \left|\frac{\theta _4}{\eta}\right|^{12} \right \}
\left (\frac{\theta _2}{\eta} \right )^4
\left \{ \overline{ \left(\frac{\theta _3}{\eta}\right )^4}
+ \overline{ \left( \frac{\theta _4}{\eta}\right) ^4 } \right \} ,
\nonumber
  \\
&Z_{2}^{(\mathrm{R} ,\mathrm{R})} (\tau , \bar \tau ) 
= -Z_{2}^{(\mathrm{NS} ,\mathrm{R})}
(\tau , \bar \tau )
=\left|\frac{\theta _2}{\eta}\right |^{20},
\nonumber
\\
%\end{align}
%\begin{align}
& Z_{1,3}^{(\mathrm{NS} ,\mathrm{NS})}
(\tau , \bar \tau ) = -Z_{1,3}^{(\mathrm{NS} ,\mathrm{R})}
(\tau , \bar \tau ) 
=\left|\frac{\theta _2}{\eta}\right |^ {12}\left\{ \left|\frac{\theta _3}{\eta }\right |^ 8
+ \left|\frac{\theta _4}{\eta }\right |^ 8\right\},
\nonumber
\\
& Z_{1,3}^{(\mathrm{R} ,\mathrm{R})} (\tau , \bar \tau ) 
= -Z_{1,3}^{(\mathrm{R} ,\mathrm{NS})}
(\tau , \bar \tau )
=\left|\frac{\theta _2}{\eta}\right |^{8}\left\{ \left|\frac{\theta _3}{\eta }\right |^{12}
+\left|\frac{\theta _4}{\eta }\right |^{12} \right\}.
 \label{part VI}  
\end{align}

This model shares  basic features with the model V as expected, 
although the mass spectrum in each sector is slightly different. 
Once again, we find 
$
Z^{\mathrm{(NS,NS)}}_{1,3}\neq -Z^{\mathrm{(R, NS)}}_{1,3} ,
$
and no massless (NS,NS) and (NS, R) states appear in the odd sectors 
as opposed to the model IV. 
%%%%%%%%%%%%%%%%%%%%%%%%%%%%%%%%%%%%%%%%%%%%%%%%%%%%%%%%%%%%%%%%%%%%%%%%%

~

%%%%%%%%%%%%%%%%%%%%%%%%%%%%%%%%%%%%%%%%%%%%%%%%%%%%%%%%%%%%%%%%%%%%%%
%%%%%%%%%%%%%%%%%%%%%%%%%%%%%%%%%%%%%%%%%%%%%%%%%%%%%%%%%%%%%%%%%%%%%%
%%%%%%%%%%%%%%%%%%%%%%%%%%%%%%%%%%%%%%%%%%%%%%%%%%%%%%%%%%%%%%%%%%%%%%

\section{Discussions about the Unitarity and Stability}
\label{Discussion}

We have studied various non-SUSY  string vacua realized as asymmetric orbifolds in section
%\ref{Construction} and 
\ref{The other non-\susic asymmetric orbifolds }.
The right-moving part of the twist operators always include the reflection $ (\-_R)^{\otimes 4}$,
% the $\mathbb Z_4$-chiral reflection,
%namely, 
with the non-trivial property 
%$(-\mathbf 1_R)^2 =(-1)^{F_R}$.
$ \left\{ (\-_R)^{\otimes 4} \right\}^2 = (-1)^{F_R}|_{\psi}$, or $ \left\{ (\-_R)^{\otimes 4} \right\}^2 = 
(-1)^{F_R}|_{\psi} \otimes (-1)^{F_R}|_{\la}$.
The torus partition functions for all of these vacua have been computed in a way 
showing manifestly the modular invariance, and they are properly $q$-expanded
as to be consistent with unitarity.
Moreover, by examining the  spectra of physical states read off from the partition functions, 
we have found all of them to be stable, namely, any tachyonic states do not appear in all the untwisted and twisted sectors.
These are likely to be common nice features of the non-SUSY string vacua of these types.
In this section, we shall try to clarify why this is the case. 
There are still various extensions or modifications of the non-SUSY vacua studied in this paper, 
and the arguments given here would be applicable to them 
rather generally.

%%%%%%%%%%%%%%%%%%%%%%%%%%%%%%%%%%%%%%%%%%%%%%%%%%%%%%%%%%%%%%%%%%%%%%%%%%%%%%%%%

We first recall some non-trivial points that are specific in our models of asymmetric orbifolds. 
First of all, 
as we emphasized several times, 
the building blocks given in subsection \ref{building} includes various phase factors. 
They are necessary to assure the modular covariance, 
and make the orbifold projections in the twisted sectors
to differ non-trivially from that for the untwisted sector. 
As we already mentioned in section 2, this is a main reason why it would not be self-evident whether our models are unitary. 

Secondly, needless to say, the absence of tachyonic instability is not  obvious for generic non-SUSY vacua. 
It is a  common feature  that non-SUSY orbifolds involving our models 
would include the `wrong GSO'  NS states in the twisted sectors, which are 
expressed typically as $\sim  \left(\frac{\theta_3}{\eta}\right)^4 + \left(\frac{\theta_4}{\eta}\right)^4 $
and would be potentially  tachyonic.

%%%%%%%%%%%%%%%%%%%%%%%%%%%%%%%%%%%%%%%%
%%%%%%%%%%%%%%%%%%%%%%%%%%%%%%%%%%%%%%%%

Now, let us start our discussions.
For the time being, we focus on the models without the shift operator
$\mathcal T_{2\pi R}|_\mathrm{base}$, which are $\bz_4$-asymmetric orbifolds. 
The partition functions are decomposed with respect to
the twisted sectors labeled by $a\in \bz_4$ in the form as, say,  
\eqn{departitionIII}.
Let us schematically denote the relevant partition functions 
as 
\begin{equation}
Z^{(s_L, s_R)}_a(\tau,\bar{\tau})  =  \frac{1}{4} \sum_{b\in \bz_4} Z^{(s_L,s_R)}_{(a,b)}(\tau,\bar{\tau}),
\label{Zab}
\end{equation} 
where $s_L$, ($s_R$) expresses the left-moving (right-moving) spin structure. 
We are only interested in the twisted sectors $a\neq 0$, since the unitarity and stability for the untwisted sector
are obvious by construction.

%%%%

%As demonstrated in subsection \ref{building},
As addressed  above, the building blocks we utilized involve
various phase factors.
Consequently, it would be useful to reinterpret the $b$-summation in \eqn{Zab} as that for  the  modular T-transformation 
$\tau \, \mapsto \, \tau+1$;
\begin{equation}
Z^{(s_L, s_R)}_a(\tau,\bar{\tau})
% & = & \frac{1}{4} \sum_{b\in \bz_4} Z^{(s_L,s_R)}_{(a,b)}(\tau,\bar{\tau})
%\nonumber \\
 =  \left\{
\begin{array}{ll}
\frac{1}{4} \left[ Z^{(s_L,s_R)}_{(a,0)}(\tau,\bar{\tau})
+ Z^{(s_L,s_R)}_{(a,0)}(\tau+1,\bar{\tau}+1)
\right],
& ~~~ (a=2), 
\\
\frac{1}{4} \sum_{\ell \in \bz_4}Z^{(s_L,s_R)}_{(a,0)}(\tau+\ell,\bar{\tau}+\ell),
& ~~~ ( a=1,3).
\end{array}
\right.
\end{equation}
Here, we made use of the modular covariance of the building blocks and the fact that 
the fermion chiral blocks $\overline{{\bf f}_{(2,b)}(\tau)}$ given in \eqn{defff(a,b)a}
vanishes for $b=1,3$ {\em for each spin structure}.
In the end, one finds that  the existence of non-trivial phase factors mentioned above 
eliminates the terms including the fractional level mismatch $h_L-h_R \not\in \bz$.
This observation makes it simpler to check the unitarity of the $q$-expansions of partition functions. 
All we have to do is just to examine 
whether the level matching terms $h_L-h_R \in \bz$ in  
the function $\frac{1}{4}Z^{(s_L,s_R)}_{(a,0)}(\tau,\bar{\tau})$ 
possess suitable $q$-expansions with positive integer coefficients\footnote
   {The factor $\frac{1}{4}$ is necessary due to the chiral GSO projection.}.
%%%
This is indeed the case for all the models given in subsection 
\ref{non-SUSY vacua}, 
as can be readily confirmed from the explicit forms of the building blocks. 
%presented in subsection \ref{building}.
We note that, actually, {\em all\/} the terms appearing in $Z^{(s_L,s_R)}_{(a,0)}(\tau,\bar{\tau})$
are $q$-expanded in this way. 

%%%%%%%%%%%%%%%%%%%%%%%%%%%%%%%%%%%%%%%%%%%%%%%%%%%%
%%%%%%%%%%%%%%%%%%%%%%%%%%%%%%%%%%%%%%%%%%%%%%%%%%%%

How about the stability of the vacua? 
Namely, we would like to understand why no tachyon appears in all the twisted sectors 
in spite of the complete SUSY violations. 
We note
\begin{itemize}
\item The leading term of each $Z^{(s_L,s_R)}_{(a,0)}(\tau,\bar{\tau})$ always has 
a non-negative conformal weight,
as is obvious from the building blocks presented
 in subsection \ref{building}.

\item Each $Z^{(s_L,s_R)}_{(a,0)}(\tau,\bar{\tau})$ includes the terms that originate 
from the  `supersymmetric' chiral blocks
$\cJ(\tau)$ or $\overline{f_{(a,0)}(\tau)} \left(\equiv \overline{{\bf f}_{(a,0)} (\tau) }\right)$ with $a=1,3$, and the leading term of 
$\cJ(\tau)$ possesses the conformal weight $\frac{1}{2}$. On the other hand,  
$\overline{{f}_{(a,0)}(\tau)}$ itself has the weight $\frac{1}{4}$, 
while the bosonic part of $(\-_R)^{\otimes 4}$ always adds the zero-point energy $\frac{1}{4}$.
%to the vacua of the sectors   $a=1,3$.

\end{itemize}
Therefore, the minimum conformal weight of the 
 {\em T-invariant terms\/} appearing in \\
 $Z^{(s_L,s_R)}_{(a,0)}(\tau,\bar{\tau})$ 
has to be equal $h = \frac{1}{2} +n$, $(n \in \bz_{\geq 0})$.
This fact is sufficient to conclude that no tachyonic states emerge
due to the observation given above.

%%%%%%%%%%%%%%%%%%%%%%%%%%%%%%%%%%%%%%%%%%%%%%%%%

We next consider the models including the shift operator 
$\mathcal T_{2\pi R}|_\mathrm{base}$. 
For our purpose 
it would be useful to {\em partially} make
the  Poisson resummation of $Z_{R, (w,m)}(\tau,\bar{\tau})$ \eqn{Rblock}
with respect to the temporal winding 
$m \in 4\bz$ and to sum up over $\any w \in a+ 4\bz$.
Then, we can obtain a schematic decomposition 
\begin{equation}
Z^{(s_L, s_R)}_a(\tau,\bar{\tau})  =  \frac{1}{4} \sum_{b\in \bz_4} 
\widetilde{Z}^{(s_L,s_R)}_{(a,b)}(\tau,\bar{\tau}),
\label{Zab2}
\end{equation} 
in place of \eqn{Zab}.
Here, $\widetilde{Z}^{(s_L,s_R)}_{(a,b)}(\tau,\bar{\tau})$ includes 
the contributions with the zero-mode part as 
\begin{equation}
 \sim q^{\frac{1}{4} \left(\frac{n}{4R} + R w\right)^2} \bar{q}^{\frac{1}{4} \left(\frac{n}{4R} - R w\right)^2},
\hspace{1cm} (n\in \bz, ~ w \in a + 4\bz),
\label{zero-mode factor}
\end{equation}
which give rise to the phase 
$e^{2\pi i \frac{nw}{4}} \equiv e^{\frac{i \pi}{2}na}$ under the T-transformation.
%%%
It is now very easy to repeat the above considerations about the unitarity and 
stability by just replacing  $Z^{(s_L,s_R)}_{(a,0)}(\tau,\bar{\tau})$
with $\widetilde{Z}^{(s_L,s_R)}_{(a,0)}(\tau,\bar{\tau})$,
leading to the same conclusion.

~

%%%%%%%%%%%%%%%%%%%%%%%%%%%%%%%%%%%%%%%%%%%%%%%%%%%%%%%%%%%%%%%%%%%%%%%%%%%%%%%%

\section*{Acknowledgments}
We would like to thank Y. Satoh for valuable discussions.
The work of Y.S in the early stage was supported by JSPS KAKENHI Grant Number 23540322
from Japan Society for the Promotion of Science (JSPS).

%%%%%%%%%%%%%%%%%%%%%%%%%%%%%%%%%%%%%%%%%%%%%%%%%%%%%%%%%%%%%%%%%%%%%%%%%%%%%%%%
%%%%%%%%%%%%%%%%%%%%%%%%%%%%%%%%%%%%%%%%%%%%%%%%%%%%%%%%%%%%%%%%%%%%%%%%%%%%%%%%
%%%%%%%%%%%%%%%%%%%%%%%%%%%%%%%%%%%%%%%%%%%%%%%%%%%%%%%%%%%%%%%%%%%%%%%%%%%%%%%%

%\newpage

~

\appendix
%\section{Poisson Resummation Formula}
%\par For a partition function $f$ on $\mathbb R$, one can use the relation
%\begin{align}
%\sum _{n\in \mathbb Z}f(n)= \sum_{k \in \mathbb Z} \hat f (k),     \label{Poisson}
%\end{align}
%where the Fourier transform $\hat f (k)$ is defined as
%\begin{align}
%\hat f (k)= \int _{\mathbb R} dx e^{-2i\pi kx}f(x).
%\end{align}

\section{Theta Functions}

In this appendix we summarize the conventions of theta functions we use in this paper \\
($q\equiv e^{2\pi i \tau }$, $y\equiv e^{2\pi i z }$ ~
$^\forall \tau \in \mathbb H ^+, ^\forall z \in \mathbb C$);
\begin{align}
&\theta _1(\tau ,z)\equiv i\sum ^{\infty}_{n=-\infty}(-1)^nq^{\frac{1}{2}(n-\frac{1}{2})^2}y^{n-\frac{1}{2}}
\equiv 2 \sin (\pi z)q^{\frac{1}{8}}\prod ^\infty _{m=1}(1-q^m)(1-yq^m)(1-y^{-1}q^m), \nonumber \\
&\theta _2(\tau ,z)\equiv \sum ^{\infty}_{n=-\infty}q^{\frac{1}{2}(n-\frac{1}{2})^2}y^{n-\frac{1}{2}}
\equiv 2 \cos (\pi z)q^{\frac{1}{8}}\prod ^\infty _{m=1}(1-q^m)(1+yq^m)(1+y^{-1}q^m), \nonumber \\
&\theta_3(\tau ,z)\equiv \sum ^{\infty}_{n=-\infty}q^{\frac{1}{2}n^2}y^n
\equiv   \prod ^\infty _{m=1}(1-q^m)(1+yq^{m-\frac{1}{2}})(1+y^{-1}q^{m-\frac{1}{2}}), \nonumber \\
&\theta_4(\tau ,z)\equiv \sum ^{\infty}_{n=-\infty}(-1)^nq^{\frac{1}{2}n^2}y^n
\equiv   \prod ^\infty _{m=1}(1-q^m)(1-yq^{m-\frac{1}{2}})(1-y^{-1}q^{m-\frac{1}{2}}) ,
\\
%\end{align}
%\begin{align}
&\eta (\tau )\equiv q^{\frac{1}{24}}\prod ^\infty _{m=1}(1-q^m).
\end{align}
%We here set $q\equiv e^{2\pi i \tau }$, $y\equiv e^{2\pi i z }\ 
%(^\forall \tau \in \mathbb H ^+, ^\forall z \in \mathbb C)$.

We often use the abbreviations;  $\theta_i \equiv \theta _i(\tau, 0)$, $\theta_i (z)  \equiv \theta _i(\tau, z)$, and 
 $\eta \equiv \eta(\tau)$.

%%%%%%%%%%%%%%%%%%%%%%%%%%%%%%%%%%%%%%%%%%%%%%%%%%%%%%%%%%%%%%%%%%%%%%%%%%%%%%%%
%\par In this paper, we use Jacobi's abstruse identity,
%\begin{align}
%\theta_3(\tau, z)^4-\theta_4(\tau, z)^4
%-\theta_2(\tau, z)^4+\theta_1(\tau, z)^4=0.
%\end{align}

%%%%%%%%%%%%%%%%%%%%%%%%%%%%%%%%%%%%%%%%%%%%%%%%%%%%%%%%%%%%%%%%%%%%%%%%%%%%%%%%%%%%%%%%
%%%%%%%%%%%%%%%%%%%%%%%%%%%%%%%%%%%%%%%%%%%%%%%%%%%%%%%%%%%%%%%%%%%%%%%%%%%%%%%%%%%%%%%%
%%%%%%%%%%%%%%%%%%%%%%%%%%%%%%%%%%%%%%%%%%%%%%%%%%%%%%%%%%%%%%%%%%%%%%%%%%%%%%%%%%%%%%%%

\newpage

\begingroup\raggedright\endgroup

\end{document}